\documentclass[11pt,a4paper]{article}
\pdfoutput=1
\usepackage{jheppub}

\usepackage[utf8]{inputenc}
\usepackage[T1]{fontenc}

\usepackage{epstopdf}
\usepackage{tabularx}
\usepackage{amsmath,amssymb, amsfonts,amsbsy,mathrsfs,graphicx, booktabs, slashed, xcolor, adjustbox}
\usepackage{latexsym}
\usepackage{hyperref}
\usepackage{bm}
\usepackage{comment}

\usepackage{tikz}
\usetikzlibrary{decorations.pathmorphing,positioning}
\usetikzlibrary{patterns}

\usepackage{pifont}

\usepackage[normalem]{ulem}

\usepackage{bbm}
\usepackage{scalerel}
\newcommand{\pmx}[1]{\begin{pmatrix} #1 \end{pmatrix}}

\def\equationautorefname~#1\null{Eq.\,(#1)\null}
\def\sectionautorefname~#1\null{Sec.\,#1\null}
\def\subsectionautorefname~#1\null{Sec.\,#1\null}
\def\figureautorefname~#1\null{Fig.\,#1\null}
\def\appendixautorefname~#1\null{App.\,#1\null}

\newcommand{\ep}{\epsilon}
\newcommand{\lam}{\lambda}

\graphicspath{{figs/}}

\usepackage{ifpdf,feynmp}
\ifpdf\fi
\makeatletter\g@addto@macro\bfseries{\boldmath}\makeatother

\newcommand{\cref}[1]{Chapter~\ref{ch:#1}}

\newcommand{\nn}{\nonumber}   
\newcommand{\nnl}{\nonumber \\}

\newcommand{\beq}{\begin{equation}} 
\newcommand{\eeq}{\end{equation}} 
\newcommand{\ba}{\begin{array}}  
\newcommand{\ea}{\end{array}} 
\newcommand{\bea}{\begin{eqnarray}}  
\newcommand{\eea}{\end{eqnarray} }  
\newcommand{\be}{\begin{eqnarray}}  
\newcommand{\ee}{\end{eqnarray} }  
\newcommand{\bal}{\begin{align}}
\newcommand{\eal}{\end{align}}   
\newcommand{\ben}{\begin{enumerate}}  
\newcommand{\een}{\end{enumerate}}  
\newcommand{\bc}{\begin{center}}
\newcommand{\ec}{\end{center}} 
\newcommand{\bt}{\begin{table}}
\newcommand{\et}{\end{table}}  
\newcommand{\btb}{\begin{tabular}}
\newcommand{\etb}{\end{tabular}}

\renewcommand{\[}{\left[}

\def\bes{\begin{equation*}}
\def\ees{\end{equation*}}
\def\bead{\begin{aligned}}
\def\eead{\end{aligned}}
\def\bmat{\left(\begin{matrix}}
\def\emat{\end{matrix}\right)}

\author[a]{Guilherme~Guedes,}
\author[b,c]{Jasper~Roosmale~Nepveu,}

\affiliation[a]{CERN, Theoretical Physics Department, Esplanade des Particules 1, Geneva 1211, Switzerland}
\affiliation[b]{Department of Physics and Center for Theoretical Physics, National Taiwan University, Taipei 10617, Taiwan}
\affiliation[c]{
Leung Center for Cosmology and Particle Astrophysics, Taipei 10617, Taiwan}

\emailAdd{guilherme.guedes@cern.ch}
\emailAdd{jasperrn@ntu.edu.tw}

\title{Two-loop renormalization of general bosonic effective field theories}

\abstract{
The renormalization of higher-dimensional operators in quantum field theory is essential for phenomenological analyses in particle physics, and plays a significant role in the study of critical phenomena.
We present a framework for renormalizing general bosonic effective field theories beyond one loop, with arbitrary gauge symmetry and scalar field content. 
In particular, we calculate the renormalization group equations in such theories up to two loops and dimension six.
When specialized to the bosonic sector of the Standard Model effective field theory (SMEFT) using simple replacement rules, our general expressions reproduce recent results from the literature. 
Due to the broad applicability of effective field theory, our general results can readily be applied to obtain the anomalous dimensions in extensions of the bosonic SMEFT and in a plethora of other theories. We also envision our results to provide useful data on the scaling dimensions of composite operators in conformal field theories without fermions.
}

\begin{document}
 \begin{flushright}
CERN-TH-2025-250
\\
\end{flushright}
\flushbottom

\makeatletter\renewcommand{\@fpheader}{\ }\makeatother
\maketitle

\section{Introduction}

The effective field theory (EFT) framework is ubiquitous in particle physics for the study of the low-energy regime of systems with scale separation. 
Within an EFT, knowledge of the renormalization group equations (RGEs) is essential to accurately relate observables measured at different energy scales and to resum large logarithms.
For this reason, there has been a tremendous effort in the literature to renormalize specific EFTs, for instance in the Standard Model effective field theory (SMEFT) at one loop and mass dimension six~\cite{Jenkins:2013wua,Jenkins:2013zja,Alonso:2013hga} and eight~\cite{Chala:2021pll,AccettulliHuber:2021uoa,DasBakshi:2022mwk,DasBakshi:2023htx,Assi:2023zid,Assi:2025fsm}; in the Low-energy effective field theory (LEFT)~\cite{Jenkins:2017dyc}; and  in extensions with axion-like particles~\cite{Chala:2020wvs,Bauer:2020jbp,DasBakshi:2023lca,Galda:2021hbr} or right-handed neutrinos~\cite{Aebischer:2024csk}.

Beyond the effort to compute the anomalous dimensions for a variety of theories, there has also been significant progress in pushing renormalization to higher perturbative orders. 
Several motivations for going beyond the one-loop level exist. From the phenomenological perspective, as precision increases on the experimental side, the theory predictions must simultaneously improve. For the SMEFT and LEFT, several works have computed the two-loop RGEs and studied their phenomenological relevance~\cite{Panico:2018hal,deVries:2019nsu,Jenkins:2023bls,Born:2024mgz,Naterop:2024cfx,Naterop:2025lzc,Ibarra:2024tpt,DiNoi:2024ajj,DiNoi:2025arz,Naterop:2025cwg,Banik:2025wpi,Zhang:2025ywe,Duhr:2025yor,Duhr:2025zqw,Haisch:2025vqj,DiNoi:2025tka,Aebischer:2025hsx,Chala:2025crd}.
In particular, important corrections to observables may be expected when the renormalization-group (RG) mixing involves Wilson coefficients that are generated at different orders by matching to UV completions~\cite{Henning:2014wua,Craig:2019wmo,Grojean:2024tcw}.
Moreover, as the state-of-the-art matching calculations for top-down SMEFT studies reach the one-loop level~\cite{Carmona:2021xtq,Fuentes-Martin:2022jrf,Guedes:2023azv,Guedes:2024vuf}, two-loop RGEs become necessary for scheme-independent final results~\cite{Ciuchini:1993fk,Ciuchini:1993ks}. 

From a theoretical point of view, multiloop renormalization results are important in the study (and verification) of non-renormalization theorems~\cite{Cheung:2015aba,Bern:2019wie} and definite signs in the RG mixing~\cite{Chala:2021wpj,Chala:2023jyx,Chala:2023xjy,Liao:2025npz}.
On the other hand, the renormalization of simpler scalar theories is useful for the development of new techniques and tools~\cite{Jenkins:2023bls,Jenkins:2023rtg,Cao:2021cdt,Cao:2023adc}.
Moreover, the multiloop anomalous dimensions in scalar EFTs in $4-2\ep$ dimensions provide valuable information outside the realm of particle physics, as they determine the spectrum of operators in Wilson--Fisher conformal field theories (CFTs)~\cite{Wilson:1971dc,Wilson:1973jj} of theoretical and experimental interest in 3d~\cite{Pelissetto:2000ek,Vasiliev2004}.

Inspired by the broad applicability of RGEs, and aiming to recycle results across a multitude of theories, it is beneficial to renormalize EFTs with general gauge symmetry and arbitrary matter content. In this approach, upon calculating the necessary integrals \emph{once and for all}, the reduction to specific theories (such as the SMEFT and its extensions) amounts to a group theory manipulation only.
For the renormalizable sector, the complete three-loop results for any theory with scalar particles, fermions and gauge bosons were obtained in Refs.~\cite{Machacek:1983fi,Machacek:1983tz,Machacek:1984zw,Mihaila:2013dta,Schienbein:2018fsw,Poole:2019kcm,Steudtner:2020tzo,Steudtner:2021fzs,Davies:2021mnc,Bednyakov:2021qxa,Steudtner:2024teg}, with partial results at four loops~\cite{Poole:2019kcm,Davies:2021mnc,Bednyakov:2021qxa,Steudtner:2025blh} and at six loops~\cite{Bednyakov:2021ojn} (building on~\cite{Kompaniets:2017yct}).
Extending this program to higher-dimensional operators in EFT, Refs.~\cite{Fonseca:2025zjb,Aebischer:2025zxg,Misiak:2025xzq} performed the one-loop renormalization of a general gauged scalar theory up to mass dimension six.
This EFT contains an arbitrary number of gauge bosons and scalar fields in unspecified representations.
Restricting to the general scalar EFT, Refs.~\cite{Henriksson:2025hwi,Henriksson:2025vyi} explored the multiloop frontier with a five-loop calculation up to dimension six, including operators with non-zero Lorentz spin, enabling a comprehensive extraction of the low-lying spectrum of operators in any $\phi^4$ Wilson--Fisher CFT.

In this paper, we initiate the renormalization of general bosonic EFTs in $4-2\ep$ dimensions beyond one loop, including scalar fields and gauge bosons, with arbitrary (unbroken) gauge and global symmetries.
Following the formalism laid out in Ref.~\cite{Fonseca:2025zjb}, we perform off-shell renormalization at two loops up to dimension six. 
For simplicity, we do not include dual field-strength tensors in the calculation, nor do we consider parameters with a positive mass dimension (i.e.~masses and $\phi^3$ couplings), but we explain how results of the latter can be obtained from our general expressions using the dummy index method~\cite{Martin:1993zk,Luo:2002ti,Schienbein:2018fsw,Bednyakov:2021ojn}.
As an important application of our results, we reproduce the RGEs in the bosonic SMEFT at two loops computed first in Ref.~\cite{Born:2024mgz} using a distinct method~\cite{Fuentes-Martin:2023ljp,Fuentes-Martin:2024agf}. 
All results are digitally available in the \texttt{GitHub} repository
\begin{center}
    \href{https://github.com/jasperrn/Bosonic-EFT-RGE}{https://github.com/jasperrn/Bosonic-EFT-RGE}\,.
\end{center}

The paper is organized as follows. 
We start by reviewing our methods and conventions in Section~\ref{sec:methods}. We then discuss our main two-loop results in Section~\ref{sec:genRes}, where we explain the notation of our auxiliary files and inspect particular anomalous dimensions of interest. 
In Section~\ref{sec:proj}, we illustrate the reduction from the general results to specific theories, focusing on the bosonic SMEFT and an $SO(N_c)$ gauged scalar theory.
We end with a brief outlook on future directions in Section~\ref{outlook}. 
We have collected relatively large expressions for some anomalous dimensions that we extracted in the $SO(N_c)$ theory in Appendix~\ref{app:RGE}.

\section{Methodology}\label{sec:methods}

In this section, we define the conventions and methodology of our computations. 
We mainly follow the formalism laid out in~\cite{Fonseca:2025zjb}, with important additions relevant to the two-loop calculations. For instance, we discuss tensor relations which are needed to simplify resulting expressions to a manageable form, and the R$^*$ method for renormalizing subdivergences and computing multiloop counterterms.
We further comment on several cross-checks that we performed.

\subsection{The general EFT Lagrangian}
The general EFT of scalars and gauge bosons has previously been introduced in Refs.~\cite{Fonseca:2025zjb,Misiak:2025xzq,Aebischer:2025zxg}. For simplicity, in this work we neglect parameters with a positive mass dimension and terms with the dual field-strength tensors ($\epsilon^{\mu\nu\rho\lambda} F^A_{\rho\lambda}$). 
However, we will show in Section~\ref{sec:posMass} how to generate the contributions from masses and $\phi^3$ couplings from our general results.
We perform the computations off shell, computing one-particle irreducible (1PI) correlation functions. Because of this, the set of operators of the general EFT must include redundant operators to absorb the full set of generated divergences, forming a so-called \emph{Green's} basis. These redundant operators (with coefficients $d_\mathcal{O}$) can be removed by field redefinitions, leading to a \emph{minimal} basis of operators (with coefficients $c_\mathcal{O}$); see Section~\ref{sec:redundant} for details on the reduction to a minimal basis.

We build the EFT Lagrangian of a multiplet of scalar fields, $\phi^a$,
with a general gauge symmetry that we assume to be an arbitrary compact Lie group with any number of Abelian and non-Abelian factors: $\mathcal{G} = \otimes_i \mathcal{G}_i$.
The general Lagrangian is thus
\begin{align}
    \mathcal{L} =& 
        - \frac{1}{4} 
        F_{\mu\nu}^AF^{\mu\nu A}
        + \frac{1}{2} D_\mu \phi^a D^\mu \phi^a
        - \frac1{4!}\lambda^{abcd}\,\phi^a\phi^b\phi^c\phi^d 
        \nnl
    &   + \frac{1}{2}d_{\phi^3D^2}^{abc} \, D^2\phi^a \phi^b \phi^c
        + \frac1{5!} c_{\phi^5}^{abcde} \, 
        \phi^a\phi^b\phi^c\phi^d\phi^e 
        + \frac{1}{2}c_{\phi F^2}^{aAB} \, 
        \phi^a F_{\mu\nu}^AF^{B \mu\nu}  \nnl 
    &   + \frac{1}{2}d_{\phi^2D^4}^{ab} D^2 \, \phi^a D^2 \phi^b
        + \frac{1}{6} d_{\phi^4D^2}^{abcd} \, D^2 \phi^a \phi^b \phi^c \phi^d
        +c_{\phi^4D^2}^{abcd} \, \phi^a \phi^b 
                D_\mu \phi^c D^\mu \phi^d \nnl 
    &   + \frac{1}{6!} c_{\phi^6}^{abcdef} \, \phi^a \phi^b \phi^c \phi^d \phi^e \phi^f
        + \frac{1}{2}d_{F^2D^2}^{AB} \, D^\mu F_{\mu\rho}^A D_\nu F^{B\nu\rho}
        + \frac{1}{6}c_{F^3}^{ABC} \, 
            F_{\mu}^{A\,\nu}
                        F_{\nu}^{B\,\rho}
                        F_{\rho}^{C\,\mu} \nnl 
    &   + \frac{1}{4}c_{\phi^2 F^2}^{abAB} \, \phi^a \phi^b F_{\mu\nu}^A F^{B\mu\nu} 
    +\frac12 d_{\phi^2 F D^2}^{abA} \,  D^{\mu} F_{\nu\mu}^{A}  \,\left(
    D^{\nu}\phi^a \phi^b -
    \phi^a D^{\nu}\phi^b \right)
    \nnl
    &+\mathcal{L}_{gf}\,,
    \label{eq:lag}
\end{align}
where the lowercase indices label all scalar fields in arbitrary representations of the gauge group, and the uppercase indices label all gauge bosons.
Throughout this paper, we will freely raise and lower these indices, and repeated indices are assumed to be summed. We absorb inverse powers of the EFT scale $\Lambda$ in the definition of the (dimensionful) couplings.

We employ the background field method~\cite{Abbott:1980hw,Abbott:1983zw}, splitting the gauge field into a quantum field, $A_q$, and a background field, $A_c$.
The gauge of the quantum fields is fixed using
\begin{equation}
    \mathcal{L}_{gf} = 
    -\frac{1}{2\xi} 
    (\widetilde{D}^\mu A_{q,\mu}^A)^2
    +\widetilde{D}_\mu \overline{c}^A
     \widetilde{D}^\mu c^A
    +f^{ABC}\widetilde{D}^\mu \overline{c}^A \, A_{q,\mu}^B \, c^C 
    \,,
\end{equation}
where $c^A$ are the ghost fields, and we work in the Feynman gauge, $\xi=1$.
The covariant derivative $\widetilde{D}_\mu$ depends only on the background field. In this construction, background gauge invariance is preserved in all steps of the calculation.

In the above theory, the gauge structure is kept generic, with field-strength tensors 
\begin{equation} 
    F^A_{\mu\nu} = \partial_\mu A^A_\nu - \partial_\nu A^A_\mu 
                    + f^{ABC} A^B_\mu A^C_\nu\,,
\end{equation}
and covariant derivatives
\begin{align}
    D_\mu \phi^a &= 
        \partial_\mu \phi^a
        -i \, T^{A}_{ab}A^A_\mu \, \phi^b \,,\nnl
    D_\mu F_{\nu\rho}^A &= 
        \partial_\mu F_{\nu\rho}^A
        +f^{ABC}A_\mu^B F_{\nu\rho}^C\,,\nnl
    \widetilde{D}_\mu A_{q,\nu}^A
     &= \partial_\mu A_{q,\nu}^A
     +f^{ABC}A_{c,\mu}^B A_{q,\nu}^C\,,\nnl
    \widetilde{D}_\mu \overline{c}^A
     &= \partial_\mu \overline{c}^A
     + f^{ABC}A_{c,\mu}^B\, \overline{c}^C
     \,,\nnl
    \widetilde{D}_\mu c^A
    &= \partial_\mu {c}^A
     + f^{ABC}A_{c,\mu}^B\,{c}^C\,,
\end{align}
where we leave the gauge coupling implicit in the definition of the tensors $f^{ABC}$ and $T^A_{ab}$\,.
We choose the generators to be antisymmetric,
\begin{align}
    T^{A}_{ab} = -T^A_{ba}\,,
\end{align}
while the structure constants $f^{ABC}$ are fully antisymmetric and satisfy the Jacobi identity,
\begin{align}\label{eq:Jacobi}
    f^{ABE}f^{CDE} + f^{ACE}f^{DBE} + f^{ADE}f^{BCE} = 0\,.
\end{align}
Finally, the structure constants are related to the generators via
\begin{align}
\label{eq:commutators}
    [T^{A},T^B] = i f^{ABC} T^C\,.
\end{align}
In the case of a theory with multiple $U(1)$ gauge groups, kinetic mixing can be generated by renormalization~\cite{delAguila:1988jz,delAguila:1995rb,Fonseca:2013jra}, even though we take the kinetic term to be diagonal in the UV, noting that it is always possible to diagonalize the kinetic term at some choice of renormalization scale $\mu$.
Following Refs.~\cite{Poole:2019kcm,Fonseca:2025zjb}, the results for kinetic mixing can be recovered in the general theory formalism by upgrading the gauge coupling to a matrix,
\begin{align}\label{eq:gaugeMatrix}
    g\,f^{ABC} \to g^{AD}f^{DBC} && g\,T^A_{ab} \to g^{AB}T^{B}_{ab}\,.
\end{align}

The symmetries of the coefficients in the general Lagrangian can be read off from the operators. For example, $\lambda^{abcd}$ is fully symmetric in its indices, while $c_{\phi^4D^2}^{abcd}$ is symmetric in the first two and the last two indices. The relabeling symmetries of all tensors are given in Table~\ref{tab:syms}.

\begin{table}[t!]
\centering
\caption{Symmetry properties of the tensors in the general theory. The tensor $c_{\phi^4D^2}$ additionally satisfies the relations of Eq.~\eqref{eq:prim}.}
\begin{tabular}{ll}
\\[-4mm]\hline
\textbf{Tensor} & \textbf{Symmetry property} \\
\hline
$\lambda^{abcd}$ & Fully symmetric\\
$d_{\phi^3D^2}^{abc}$        & Symmetric in the last two indices \\
$c_{\phi^5}^{abcde}$         & Fully symmetric \\
$c_{\phi F^2}^{aAB}$         & Symmetric in the last two indices \\
$d_{\phi^2D^4}^{ab}$       & Symmetric \\
$d_{\phi^4D^2}^{abcd}$       & Symmetric in the last three indices \\
$c_{\phi^4D^2}^{abcd}$       & Symmetric in the first two and in the last two indices \\
$c_{\phi^6}^{abcdef}$        & Fully symmetric \\
$d_{F^2D^2}^{AB}$            & Symmetric \\
$c_{F^3}^{ABC}$              & Fully antisymmetric \\
$c_{\phi^2F^2}^{abAB}$       & Symmetric in the first two and in the last two indices \\
$d_{\phi^2FD^2}^{abA}$       & Antisymmetric in the first two indices \\
\hline
\end{tabular}
\label{tab:syms}
\end{table}

\noindent
Besides the relabeling symmetry that follows directly from the form of the 
$\phi^4D^2$ operator in the minimal basis, the tensor $c_{\phi^4D^2}$ satisfies the additional relations
\begin{align}\label{eq:prim}
    c^{abcd}_{\phi^4D^2} = c^{cdab}_{\phi^4D^2}\,, 
    && c^{abcd}_{\phi^4D^2} + c^{acdb}_{\phi^4D^2} + c^{adbc}_{\phi^4D^2} = 0\,,
\end{align}
which serve to remove the ambiguity between $d_{\phi^4D^2}^{abcd}$ and $c_{\phi^4D^2}^{abcd}$ under integration by parts (i.e.~adding total derivatives). Using \eqref{eq:prim}, the corresponding operator can be rewritten as
\begin{align}
    c_{\phi^4D^2}^{abcd} \phi^a \phi^b D_\mu \phi^c D^\mu \phi^c
    = \tfrac{1}{3} c^{abcd}_{\phi^4D^2}
    \Big( D_\mu\phi^a D^\mu\phi^b \phi^c \phi^d 
            + \phi^a \phi^b D_\mu \phi^c D^\mu \phi^d 
            -\tfrac12 D_\mu\phi^a\phi^b D^\mu\phi^c \phi^d \nnl
            -\tfrac12 D_\mu\phi^a\phi^b \phi^c D^\mu\phi^d
            -\tfrac12 \phi^a D_\mu\phi^b D^\mu\phi^c \phi^d
            -\tfrac12 \phi^a D_\mu\phi^b \phi^c D^\mu\phi^d
    \Big)\,.
\end{align}
which matches the convention of~\cite{Fonseca:2025zjb}. 
We will explain the origin of the condition~\eqref{eq:prim} in more detail in the next subsection.

\subsection{Constraints from integration by parts and field redefinitions}
\label{s:primop}

To avoid ambiguities in the calculation and in the presentation of results, it is important to avoid redundant parameters in the Lagrangian. 
This applies both to the full Green's basis and the minimal basis. The former contains all operators that contribute independently to off-shell correlation functions, while operators in the minimal basis contribute independently to S-matrix elements. 
We have split the Lagrangian~\eqref{eq:lag} into operators parameterized by tensors $d_{\mathcal{O}}$ and $c_\mathcal{O}$. The Green's basis is spanned by all operators, while a minimal basis is given by all operators with couplings $c_\mathcal{O}$.

In the general theory, ambiguities are avoided by imposing symmetries on the tensors. While the symmetries of most operators follow straightforwardly from their field and derivative content, this is not the case for the $\phi^2FD^2$ and for one of the $\phi^4D^2$ operators.
The antisymmetry of the first two indices of $d_{\phi^2FD^2}^{abA}$ in the operator
\begin{equation}
    d_{\phi^2FD^2}^{abA} D^\mu F_{\mu\nu}^A D^\nu \phi^a \phi^b = 
    \frac12 d_{\phi^2FD^2}^{abA} D^\mu F_{\mu\nu}^A (D^\nu \phi^a \phi^b -\phi^a  D^\nu  \phi^b)
\end{equation}
follows from the fact that the symmetric components are redundant,
\begin{align}
    \frac12 D^\mu F_{\mu\nu}^A D^\nu \phi^a \phi^b + (a \leftrightarrow b)
    &= \frac12 D^\mu F_{\mu\nu}^A D^\nu (\phi^a \phi^b)
    \stackrel{ibp}{=}
     \frac12 D^\nu D^\mu F_{\mu\nu}^A  \phi^a \phi^b
     =0 \,.
\end{align}
In the second equality, we indicate by $ibp$ (integration by parts) that we ignore a total derivative operator,
which does not contribute to off-shell correlation functions nor to S-matrix elements.
The final equality follows from the fact that $[D^\mu,D^\nu]F_{\mu\nu}^A = f^{ABC}F^B_{\mu\nu}F^{C\mu\nu} = 0$.

The constraints on $c_{\phi^4D^2}$, given in \eqref{eq:prim}, take a more complicated form, which we will discuss in more detail below. We will take guidance from the Hilbert series~\cite{Henning:2015alf,Henning:2017fpj} to count the number of independent parameters of the tensors in the pure scalar case, both in the Green's basis and in the minimal basis.
The discussion for the $\phi^4D^2$-type operators is meant to generalize to more complicated operators at higher mass dimensions. (See \cite[App.~B]{Henriksson:2025vyi} for the dimension-eight operators in the minimal basis in a general theory with only scalar fields.)

\paragraph{\boldmath$\phi^4\partial^2$ operators in the Green's basis.}

The Hilbert series in a general theory of $n$ scalar fields in the Green's basis at mass dimension six is given by
\begin{align}\label{HGreens}
    H^{(6,\text{Green's})} = 
    \binom{n+1}{2}\phi^2\partial^4
    + \binom{n+1}{2}^2 \phi^4 \partial^2
    + \binom{n+5}{6} \phi^6\,,
\end{align}
which lists the number of independent parameters per type of operator, where the terms in parenthesis correspond to the binomial coefficient, $\binom{n}{k}$, which is nonzero for $n\geq k$. For $\phi^2\partial^4$ and $\phi^6$ these are simply the number of parameters in two- and six-index symmetric tensors, respectively. Let us unpack the number of independent $\phi^4\partial^2$ operators more slowly. 

Any $\phi^4\partial^2$ operator can be written as
\begin{align}
    \mathcal{O}_{\phi^4\partial^2} = 
    c_1^{abcd} \, \phi^a \phi^b \phi^c \partial^2\phi^d
    + c_2^{abcd} \, \phi^a \phi^b \partial_\mu \phi^c \partial^\mu \phi^d \,.
    \label{eq:Greens1}
\end{align}
In the Green's basis, we can use integration by parts to relate the first term to the second,
\begin{align}
    &\mathcal{O}_{\phi^4\partial^2} \stackrel{ibp}{=} 
     c_3^{abcd} \, \phi^a \phi^b \partial_\mu \phi^c \partial^\mu \phi^d\,, \quad \text{with} \quad
    %
     c_3^{abcd} = 
    -c_1^{abcd} - c_1^{acbd} - c_1^{bcad} + c_2^{abcd}\,.
    \label{eq:Greens2}
\end{align}
Here $c_3^{abcd}$ is symmetric in the first two and in the last two indices, which means that the number of independent parameters is $\binom{n+1}{2}^2$, consistent with \eqref{HGreens}.

\paragraph{\boldmath$\phi^4\partial^2$ operators in the minimal basis.}

The number of operators in the minimal basis is given by
\begin{align}\label{eq:Hmin}
    H^{(6,\text{minimal})} = 
    \frac{n}{2}\binom{n+1}{3} \phi^4 \partial^2
    + \binom{n+5}{6} \phi^6\,.
\end{align}
In particular, there is no $\phi^4 \partial^2$ operator in a single-scalar theory and there is a single operator for $n=2$.
One choice of minimal basis can be obtained from the Green's basis by removing all operators proportional to $\partial^2\phi^a$. 
Indeed, there are 
$n\binom{n+2}{3}$ operators of the form $\phi^a\phi^b\phi^c \partial^2\phi^d$, leading to the correct counting:
\begin{equation}
    \binom{n+1}{2}^2 - \binom{n+2}{3}
    = \frac{n}{2}\binom{n+1}{3}\,.
\end{equation}

At the level of the operators, our aim is to impose tensor relations on the tensors in~\eqref{eq:Greens1} to avoid ambiguities under integration by parts. 
For instance, if $c_2^{abcd}$ is fully permutation invariant, \eqref{eq:Greens1} can be rewritten as
\begin{align}
    \mathcal{O}_{\phi^4\partial^2} \stackrel{ibp}{=} 
    \left(c_1^{abcd}-\frac13 c_2^{abcd}\right) \phi^a \phi^b \phi^c \partial^2\phi^d \,,
    \label{eq:Greens3}
\end{align}
which shows that there is an ambiguity between the parameters of $c_1$ and $c_2$ unless we impose additional constraints on the tensors. One choice of minimal basis of EFT operators is given by the conformal primary operators~\cite{Henning:2015alf,Henning:2017fpj}, which satisfy $[K_\mu, \mathcal{O}(0)]=0$, where $K_\mu$ is the generator of special conformal transformations. This leads to the conditions \eqref{eq:prim} previously derived in Refs.~\cite{Henriksson:2025hwi,Henriksson:2025vyi} in this way.
These constraints remove the total derivatives from the parameters, and are such that $c_{\phi^4D^2}^{abcd}$ corresponds to the operators in the minimal basis.%
    \footnote{It can explicitly be checked that $c_{\phi^4D^2}^{abcd}$ contains the correct number of independent parameters upon imposing the constraints \eqref{eq:prim}. 
    There is no solution to the constraints when all indices are the same, e.g.~$c_{\phi^4D^2}^{1111}=0$.
    When two indices are different, there is one independent ordering, e.g.~$c_{\phi^4D^2}^{1122}$, leading to $\binom{n}{2}$ independent operators. 
    For operators with three and four distinct fields, there are 
    $3\binom{n}{3}$ and
     $2\binom{n}{4}$ independent parameters, respectively. The sum of these reproduces the number of $\phi^4\partial^2$ operators in~\eqref{eq:Hmin}.
    }

\subsection{Constraints from gauge invariance}
\label{sec:gaugerelations}

In the previous section, we described the conditions on the tensors that arise from integration by parts and field redefinitions. There are further constraints on the tensors to restrict to gauge-invariant operators. 
For example, an infinitesimal gauge transformation, $\alpha^C(x)$, on the $\phi F^2$-type operators leads to
\begin{equation}
    \alpha^C(x) 
    \left(iT^C_{ab} c^{bAB}_{\phi F^2} - f^{ACD} c^{aDB}_{\phi F^2} - f^{BCD} c^{aAD}_{\phi F^2}
    \right)\phi^aF^A_{\mu\nu}F^{B\mu\nu} + O(\alpha^2)
    =0 \,.
\end{equation}
For singlet scalars $\phi^a$, the generator vanishes, $T^A_{ab} = 0$. 
In this case, the constraint can be satisfied by $c_{\phi F^2}^{aAB}=\tilde c_{\phi F^2}^a g^{AB}$, where $g^{AB}$ is the gauge coupling matrix as in \eqref{eq:gaugeMatrix}.
The remaining conditions on the coupling constant tensors in the minimal basis are
\begin{align}
T^A_{ae} \lambda^{ebcd}
+T^A_{be} \lambda^{aecd}
+T^A_{ce} \lambda^{abed}
+T^A_{de} \lambda^{abce} &= 0 \,,\nn\\[1mm]
T^A_{af} c_{\phi^5}^{fbcde}
+T^A_{bf} c_{\phi^5}^{afcde}
+T^A_{cf} c_{\phi^5}^{abfde}
+T^A_{df} c_{\phi^5}^{abcfe}
+T^A_{ef} c_{\phi^5}^{abcdf} &= 0 \,,\nn\\[1mm]
i f^{ACD} c_{\phi F^2}^{aDB} 
+ i f^{BCI} c_{\phi F^2}^{aAD} 
+ T^C_{ab} c_{\phi F^2}^{bAB} &=0\,,\nn\\[1mm]
T^A_{ae} c_{\phi^4D^2}^{ebcd}
+T^A_{be} c_{\phi^4D^2}^{aecd}
+T^A_{ce} c_{\phi^4D^2}^{abed}
+T^A_{de} c_{\phi^4D^2}^{abce} &= 0 \,,\nn\\[1mm]
T^A_{ag} c_{\phi^6}^{gbcdef}
+T^A_{bg} c_{\phi^6}^{agcdef}
+T^A_{cg} c_{\phi^6}^{abgdef}
+T^A_{dg} c_{\phi^6}^{abcgef}
+T^A_{eg} c_{\phi^6}^{abcdgf} 
+T^A_{fg} c_{\phi^6}^{abcdeg} &= 0 \,,\nn\\[1mm]
f^{ADE}c_{F^3}^{EBC}
+f^{BDE}c_{F^3}^{AEC}
+f^{CDE}c_{F^3}^{ABE}&=0\,,\nn\\[1mm]
i f^{ACD} c_{\phi^2F^2}^{abDB} 
+ i f^{BCD} c_{\phi^2F^2}^{abAD} 
+ T^C_{ac} c_{\phi^2F^2}^{cbAB}
+ T^C_{bc} c_{\phi^2F^2}^{acAB} &=0\,.
\label{eq:gaugerelations}
\end{align}
Besides these relations, the tensors $f^{ABC}$ satisfy Jacobi identities \eqref{eq:Jacobi} and the generators are related to the structure constants via Eq.~\eqref{eq:commutators}.

In principle, it is not necessary to impose the relations of the form \eqref{eq:gaugerelations} before projecting the results onto a specific theory, because they will necessarily be satisfied by the tensors of the chosen gauge theory. 
Nevertheless, in our general results, we simplified the computed anomalous dimensions using the constraints from gauge invariance to reduce the size of the expressions and to enable certain cross-checks and consistency checks, as we will discuss in Section~\ref{sec:simplification}.

\subsection{Definitions of the RGEs}
\label{sec:computation}

Having discussed the operator content of the general theory, we now define the beta functions and how to compute them in terms of the counterterms. We work in the (modified) minimal subtraction scheme of dimensional regularization in $d=4-2\ep$ dimensions. 

We define beta functions and the anomalous dimension of the fields to be 
\begin{align}\label{eq:ADfields}
    \frac{d\,c}{d\log\mu} = \beta_c \,, &&
    \frac{d\,\phi^a}{d\log\mu} = \gamma_\phi^{ab}\phi^b\,,&&
    \frac{d\,A_\nu^B}{d\log\mu} = \gamma_A^{BC}A_\nu^C\,,
\end{align}
for any coupling $c$ (leaving the indices implicit), where $\mu$ is the renormalization scale.
Note that our convention for $\gamma_\phi$ and $\gamma_A$ has an overall minus sign with respect to more standard conventions.
These RGEs can be computed after renormalizing the theory, using the fact that bare couplings $c_0$ and bare fields are $\mu$-independent,
\begin{align}
    \frac{d (\mu^{(n_c-2)\epsilon} \, c_0) }{d\log\mu} = 0 \,,
    &&
    \frac{d (Z_\phi^{ab} \phi^a) }{d\log\mu} = 0 \,,
    &&
    \frac{d (Z_A^{BC} A_\nu^C) }{d\log\mu} = 0 \,,
\end{align}
where we define the couplings $c_0$ in terms of the counterterms below, and $n_c$ is the number of fields in the operator that $c_0$ multiplies (treating covariant derivatives as normal derivatives for this counting). The counterterms $Z_\phi$ and $Z_A$ are obtained by renormalizing the kinetic terms and they depend on the couplings from the dimension-four Lagrangian only.
More precisely, 
\begin{align}
\mathcal{Z} \big( \Gamma_{2,0}(p_1,p_2) \big) &= 
\mathcal{Z}\!\left(    
    \vcenter{\hbox{\begin{tikzpicture}[baseline=(B.center), scale=0.9]
  \tikzset{
    blob/.style={draw, thick, circle, minimum size=18pt, inner sep=0pt, fill=gray!10,
    pattern=north east lines,      
    pattern color=black!40},
    blobb/.style={draw, thick, circle, minimum size=18pt, inner sep=0pt, fill=gray!10}
  }
  \coordinate (L) at (-1.0,0);
  \coordinate (R) at ( 1.0,0);
  \draw (L) -- (R) node[midway, above] {};
  \node[blobb] (B) at (0,0) {};
  \node[blob] (B) at (0,0) {};
\end{tikzpicture}
}} \right) = i(Z_\phi)^2 \, (-p_1\cdot p_2) \nn\\
\mathcal{Z} \big( \Gamma_{0,2}(p_1,p_2) \big) &= 
\mathcal{Z}\!\left(    
    \vcenter{\hbox{\begin{tikzpicture}[baseline=(B.center), scale=0.9]
  \tikzset{
    photon/.style={decorate, decoration={snake, amplitude=1.5pt, segment length=5.0pt}, line width=1pt},
    blob/.style={draw, thick, circle, minimum size=18pt, inner sep=0pt, fill=gray!10,
    pattern=north east lines,      
    pattern color=black!40},
    blobb/.style={draw, thick, circle, minimum size=18pt, inner sep=0pt, fill=gray!10}
  }
  \coordinate (L) at (-1.0,0);
  \coordinate (R) at ( 1.0,0);
  \draw[photon] (L) -- (R) node[midway, above] {};
  \node[blobb] (B) at (0,0) {};
  \node[blob] (B) at (0,0) {};
\end{tikzpicture}
}} \right) = i(Z_A)^2 \, (p_1\cdot p_2 \,\eta^{\mu_1 \mu_2} - p_1^{\mu_1} p_2^{\mu_2})
\end{align}
where we leave flavor and gauge indices implicit. Here $\mathcal{Z}$ extracts the counterterm as minus the overall UV divergence of the 1PI correlator, which we define as
\begin{align}
    \Gamma_{n,m} = \int \left(
    \prod_{i=1}^{n+m} d^dx_i \, e^{ip_i\cdot x_i}
    \right)
    \langle0|T \{ \phi_{x_1}\cdots\phi_{x_n}
    A^{\mu_1}_{x_{n+1}}\cdots A^{\mu_{m}}_{x_{n+m}}\} |0\rangle_\text{1PI} \,,
\end{align}
where $\phi_x = \phi(x)$ and similarly for the gauge boson fields.

At the order that we work with, the bare couplings at different mass dimensions take distinct forms. At dimension four and five, we have
\begin{align}
    \lambda_0^{abcd} = Z_4^{abcd}\,,
    &&
    c_0^{\mathcal{O}_5} = \sum_{\bar{\mathcal{O}}_5}
    Z_5^{\mathcal{O}_5,\bar {\mathcal{O}}_5}
    c^{\bar{\mathcal{O}}_5}\,,
\end{align}
where we use superscript $\mathcal{O}_5$ to denote both the respective dimension-five operators and their indices.
In our calculation, $Z_4$ and $Z_5$ depend on $\lambda$ and the gauge couplings only.%
    \footnote{We remind the reader that we do not explicitly include parameters with positive mass dimension, such as masses or $\phi^3$ couplings, in the Lagrangian. Nevertheless, we describe how these contributions can be obtained from our general results in Section~\ref{sec:posMass}.}
They can (schematically) be obtained from the correlation functions as
\begin{align}
    \mathcal{Z}(\Gamma_{4,0}) &= i\,(Z_\phi)^4\, Z_4\,, \nn\\ 
    \mathcal{Z}(\Gamma_{n,m}[\tilde{\mathcal{O}}_5])&= 
    (Z_\phi)^{n} (Z_A)^m \sum_{\mathcal{O}_5} Z_5^{\mathcal{O}_5,\tilde{\mathcal{O}}_5} \, \text{FR}(\mathcal{O}_5)\,,
    \label{eq:dim5CT}
\end{align}
where $\Gamma_{n,m}[\mathcal{O}]$ denotes the 1PI correlator with an extra insertion of operator $\mathcal{O}$. The sum in \eqref{eq:dim5CT} runs over all operators in the Green's basis, and $\text{FR}(\mathcal{O})$ denotes the Feynman rule of operator~$\mathcal{O}$.

At dimension six, we also include contributions from two insertions of dimension-five operators,
\begin{align}
    c_0^{\mathcal{O}_6} = 
    \sum_{\bar{\mathcal{O}}_6}
    Z_6^{\mathcal{O}_6,\bar {\mathcal{O}}_6}
    c^{\bar{\mathcal{O}}_6} + 
    \sum_{\bar{\mathcal{O}}_5 \tilde{\mathcal{O}}_5}
    Z_{5,5}^{\mathcal{O}_6,\bar {\mathcal{O}}_5 \tilde{\mathcal{O}}_5}    c^{\bar{\mathcal{O}}_5}c^{\tilde{\mathcal{O}}_5}\,,
\end{align}
where the counterterms are obtained as in \eqref{eq:dim5CT}, computing correlation functions with two insertions of dimension-five operators for $Z_{5,5}^{\mathcal{O}_6,\bar {\mathcal{O}}_5 \tilde{\mathcal{O}}_5}$.

In terms of these counterterms, the beta functions and anomalous dimensions of the fields can then be computed to be
\begin{align}
    \gamma_\phi^{ab} &= L \, Z_\phi^{ab}\Big|_{\ep^{-1}} \,,
    \qquad
    \gamma_A^{BC} =  L \, Z_A^{BC}\Big|_{\ep^{-1}} \,,\nn\\[2mm]
    \beta_\lambda^{abcd} &= -2\ep \lambda^{abcd} 
    +2 L 
    \left(Z_4^{abcd}
        \Big|_{\ep^{-1}}\right) 
    \,,\nn\\[2mm]
    \beta_5^{\mathcal{O}_5} &=
    -(n^{\mathcal{O}_5}-2) \ep 
    +2 L 
    \sum_{\bar{\mathcal{O}}_5}
    \left( Z_5^{\mathcal{O}_5,\bar {\mathcal{O}}_5} \Big|_{\ep^{-1}} \right)
    c^{\bar{\mathcal{O}}_5}\,,\nn\\[2mm]
    \beta_6^{\mathcal{O}_6} &=
    -(n^{\mathcal{O}_6}-2) \ep 
    + 2L
    \sum_{\bar{\mathcal{O}}_6}
    \left( Z_6^{\mathcal{O}_6,\bar {\mathcal{O}}_6}
    \Big|_{\ep^{-1}}
    \right)
    c^{\bar{\mathcal{O}}_6} + 
    2L\sum_{\bar{\mathcal{O}}_5 \tilde{\mathcal{O}}_5}
   \left( Z_{5,5}^{\mathcal{O}_6,\bar {\mathcal{O}}_5 \tilde{\mathcal{O}}_5}   
    \Big|_{\ep^{-1}}
    \right)
    c^{\bar{\mathcal{O}}_5}c^{\tilde{\mathcal{O}}_5}\,, \label{eq:RGEfrompole}
\end{align}
where $L$ takes the value 1 when it multiplies a one-loop expression and 2 at two loops, etc., and we restrict to the $1/\ep$ poles of the counterterms. Since our calculation includes correlation functions with insertions of all operators of the Green's basis, $\beta_5$ and $\beta_6$ can be evaluated for the full Green's basis. 
Alternatively, one can first reduce the results to a minimal basis by a field redefinition, in which case the couplings of the redundant operators ($d_\mathcal{O}$) can be set to zero; see Section~\ref{sec:redundant}.

Even though we extract the beta functions and anomalous dimensions from the $1/\ep$ poles of the counterterms in \eqref{eq:RGEfrompole}, we note that the RGEs can also be computed from the full counterterms, including all poles. Consistency with \eqref{eq:RGEfrompole} leads to conditions on the higher-order poles~\cite{tHooft:1973mfk}. In fact, the above expression assumes that the RGE's are finite, which is not always the case; see Section~\ref{sec:cross}. Equation~\eqref{eq:RGEfrompole} then only captures the finite part.

In the background field method, the beta function of the gauge couplings can be obtained from the anomalous dimension of the field,
\begin{align}\label{betagamma}
    \beta_{g}^{BC} = -\ep g^{BC} - \gamma_A^{BC}\,,
\end{align}
since 
$g^{BC} A_\mu^C$ is not renormalized.

\subsection{Counterterm calculation: the R$^*$ method}

To obtain the counterterms from which the RGEs are determined, we compute the UV divergences of off-shell one-particle-irreducible (1PI) correlation functions using the R$^*$ method~\cite{Vladimirov:1979zm,Chetyrkin:1982nn,Chetyrkin:1984xa,Smirnov:1985yck,Larin:2002sc,Kleinert:2001hn,Batkovich:2014rka,Chetyrkin:2017ppe,Beekveldt:2020kzk}.
One of the main features of this method is the approach due to Bogoliubov--Parasiuk--Hepp--Zimmermann (BPHZ)~\cite{Bogoliubov:1957gp,Hepp:1966eg,Zimmermann:1969jj}
for recursively subtracting subdivergences from Feynman integrals; see also~\cite{Collins:1984xc} for a textbook explanation. 
In this approach, called \emph{local} renormalization, one does not explicitly add lower-loop counterterm diagrams to cancel subdivergences, but divergent subdiagrams are automatically identified and subtracted.
As a result, the remaining divergence is a local (i.e.~polynomial) function of the momenta. 

Another important advantage of the R$^*$ method is the use of 
infrared rearrangement~\cite{Vladimirov:1979zm} to simplify an $L$-loop multiscale integral into a product of massless two-point diagrams of at most $L-1$ loops. 
In our implementation, this is achieved by first Taylor expanding the loop integral in combination with the expansion-by-subgraph~\cite{Smirnov:1990rz, Smirnov:1994tg,Smirnov:2002pj,Chakraborty2023,Chakraborty:2024uzz}.
This leads to a sum over logarithmically divergent integrals, with a single off-shell momentum routed arbitrarily through the diagrams. With a suitable choice of routing, namely in and out of adjacent vertices, the integral can then be factorized into lower-loop integrals~\cite{Chetyrkin:1982nn}.

The implementation of the R$^*$ method that we employ has previously been applied in a multitude of contexts~\cite{Herzog:2017ohr,Herzog:2017bjx,Herzog:2017dtz,Herzog:2018kwj,deVries:2019nsu,Cao:2021cdt,Cao:2023adc,Henriksson:2025hwi,Henriksson:2025vyi}, and we refer to these references for more details; see also Refs.~\cite{Naterop:2024cfx,Born:2024mgz} for other recent applications of R$^*$ in EFT.
Our framework is based on \texttt{Maple}~\cite{maple} and \texttt{Form}~\cite{Vermaseren:2000nd,Ruijl:2017dtg}, and we rely on the programs 
\texttt{Opiter}~\cite{Goode:2024mci,Goode:2024cfy} and
\texttt{Forcer}~\cite{Ruijl:2017cxj,Baikov:2010hf,Lee:2011jt} for 
the tensor reduction and the integral reduction, respectively.
We use the diagram generator of Refs.~\cite{Kaneko:1994fd,Kaneko:2017wzd} in \texttt{Form}.

\subsection{Simplification of tensor expressions}
\label{sec:simplification}

After calculating the counterterms, we simplify the resulting expressions by identifying terms that are equivalent under 
(\emph{i})~the symmetries of the tensors,
(\emph{ii})~dummy index relabeling, and
(\emph{iii})~the constraints on tensor products that follow from gauge invariance. 
These simplifications are crucial for comparing the results to existing calculations in the literature, as well as for internal cross-checks, which we describe in the next section.

As for (\emph{i}), the symmetry properties of functions can be specified when they are defined in \texttt{Form}. 
Importantly, we also reduced the size of the resulting expressions by equating terms that are related by permutations of the external indices. For example, the beta function of $c_{\phi^6}^{abcdef}$ is fully permutation invariant, which we exploited by freely relabeling the open indices. The full expression can then be obtained by summing all permutations normalized by $1/6!$. 
In the expressions that we make publicly available, we keep track of the assumed permutation symmetries by contracting the open indices of the beta functions by auxiliary tensors with definite symmetries. This will be exemplified in more detail in Section~\ref{sec:genRes}. 

To remove the redundancy associated with point (\emph{ii}), we identify products of tensors with elements in a set of non-isomorphic diagrams, generated with \texttt{qgraf}~\cite{Nogueira:1991ex}. In this way, any two terms that are related by relabeling dummy indices are mapped to the same diagram, thus removing this ambiguity.

To address the relations between tensors that follow from gauge invariance (\emph{iii}), 
we follow the algorithm described in Ref.~\cite{Poole:2019kcm},
which can be applied in EFTs and at any loop order. 
We start from all tensor structures associated to the generated (non-isomorphic) diagrams. Importantly, this set of diagrams must include non-1PI and tadpole diagrams, because these structures can be generated by the gauge relations. We then rewrite each inequivalent tensor expression applying the gauge transformations it must respect, such as \eqref{eq:gaugerelations}, with free parameters. By choosing different numerical values for these parameters, we generate a system of equations that involves the set of tensor expressions.%
    \footnote{In this method of simplification, we do not guarantee that all tensor relations are discovered. It might be that some relations are missed because the parameters are not evaluated for enough different numerical values. We performed all simplifications that were necessary for the internal consistency checks described in Section~\ref{sec:cross}.}
The solution of this system takes the form of linear relations, which we then use to simplify our expressions. 
In our current implementation, we do not predefine the basis of tensors that we keep in our simplified expressions.%
    \footnote{In addition to our own implementation, on various occasions we also benefited from the \texttt{Mathematica} package \texttt{SimTeEx}~\cite{Fonseca:2024rcg} to simplify tensor expressions.}

This identification of equivalent tensors greatly reduces the size of the expressions. For example, the two-loop counterterm for the dimension-six $\phi^2 F^2$ operators simplifies from 
6320 to only 129 terms. (This counting concerns a single permutation of the open indices, after following (\emph{i}) to partially simplify the expression.)
On the other hand, our current implementation would require further optimization before extending to higher loop orders.%
    \footnote{In fact, we did not simplify the expressions for the $\phi^6$ operator at two loops. The provided expressions are thus expected to be correct, yet in a redundant form.}
The pattern recognition that is required for the simplification procedure that we describe here is computationally expensive.

\subsection{Cross-checks and consistency checks}\label{sec:cross}

We have cross-checked the implemented method and the results in several ways, which we describe in this section.
When possible, we compared the computed anomalous dimensions with existing results. 
At dimension four, we are in full agreement with the two-loop results of 
Refs.~\cite{Machacek:1983tz, Machacek:1984zw}. 
At dimension five and dimension six in the pure-scalar limit (and for single insertions of the EFT operators), we compared our two-loop results with those of Refs.~\cite{Henriksson:2025hwi,Henriksson:2025vyi}.
The general bosonic EFT has recently been renormalized at one loop~\cite{Fonseca:2025zjb,Aebischer:2025zxg,Misiak:2025xzq}; we cross-checked our results at the Green's basis level with Ref.~\cite{Fonseca:2025zjb}, and after reducing to the minimal basis with Refs.~\cite{Fonseca:2025zjb,Aebischer:2025zxg}. 
At two loops and dimension six, we specified our general results to the particle content and gauge symmetry of the (bosonic) SMEFT, reproducing the results of~\cite{Born:2024mgz}; see Section~\ref{sec:SMEFT} for more details.
Since the bosonic SMEFT does not involve dimension-five operators, we have also specified our results to an $SO(N_c)$ gauged scalar theory coupled to $N_f$ singlet scalars with arbitrary interactions (Section~\ref{sec:scalarQCD}).
The additional consistency checks in this calculation further corroborate our results.

Our implementation also allows for a number of internal consistency checks. Firstly, the R$^*$ method automatically subtracts the UV subdivergences from the diagrams, leaving an overall divergence which is polynomial in the momenta. The absence of logarithms is thus a necessary condition on the computed counterterms. In fact, the five-loop calculation of Refs.~\cite{Henriksson:2025hwi,Henriksson:2025vyi} previously validated the same implementation for EFT renormalization that we employed in the current work.

As another check on the counterterms, we verified that they are consistent with gauge invariance, which is necessary in the background field method.%
    \footnote{Since subdivergences are automatically subtracted in the employed R$^*$ method, we do not need to compute counterterms to gauge-variant operators, which would otherwise be necessary to cancel subdivergences.}
For instance, gauge invariance requires that the two-point function of background gauge bosons at dimension four has the kinematic structure
\begin{equation}
    \vcenter{\hbox{\begin{tikzpicture}[baseline=(B.center), scale=0.9]
  \tikzset{
    photon/.style={decorate, decoration={snake, amplitude=1.5pt, segment length=5.0pt}, line width=1pt},
    blob/.style={draw, thick, circle, minimum size=18pt, inner sep=0pt, fill=gray!10,
    pattern=north east lines,      
    pattern color=black!40},
    blobb/.style={draw, thick, circle, minimum size=18pt, inner sep=0pt, fill=gray!10}
  }
  \coordinate (L) at (-1.0,0);
  \coordinate (R) at ( 1.0,0);
  \draw[photon] (L) -- (R) node[midway, above] {};
  \node[blobb] (B) at (0,0) {};
  \node[blob] (B) at (0,0) {};
\end{tikzpicture}
}}\! \propto \, \left( p_1\cdot p_2 \, \eta^{\mu\nu} - p_{1}^{\mu} p_{2}^{\nu} \right).
\end{equation}
The confirmation of this structure and similar structures in other calculations requires mapping the tensor expressions into a basis, as described in Section~\ref{sec:simplification}.

There are additional consistency conditions on the counterterms that arise from the fact that anomalous dimensions (of physical quantities) must be finite as $\ep\to0$~\cite{tHooft:1973mfk}. This constrains the coefficient of the $1/\ep^2$ poles at two loops in terms of the one-loop $1/\ep$ poles. Since the two-loop anomalous dimension is determined from the $1/\ep$ pole, this condition does not in fact validate the full results. However, it provides a strong internal consistency check on the choice of operator basis, the associated Feynman rules, and the diagram evaluation. 
As we perform the calculation in the full Green's basis, including insertions of redundant operators, the consistency conditions on the $1/\ep^n$ poles can be checked before performing field redefinitions. However, we have gauged fixed $\xi=1$ at the start, which violates the consistency conditions on the counterterms which are $\xi$-dependent, such as those of the gauged scalar fields.
This then leads to remaining $1/\ep$ poles in the anomalous dimensions.%
    \footnote{A similar observation was made in Ref.~\cite{Manohar:2024xbh}, which pointed out that setting redundant parameters to zero too early in the calculation leads to infinite anomalous dimensions of the fields, even in a pure scalar theory.}

Let us observe this explicitly in the general theory. 
Using the convention of Eq.~\eqref{eq:ADfields}, the anomalous dimension of the gauge field is~\cite{Machacek:1984zw} 
\begin{align}
\label{eq:gammag}
    \gamma_A^{BC} 
    &= g^2 \,l \left[\frac{11}{3}f^{BDE}f^{CDE}
            -\frac{1}{6} T^B_{ab}T^{C}_{ba}\right] \nn\\
                &\quad 
                +g^4 \, l^2 \left[ 
                \frac{34}{3} f^{BEF}f^{DEF}f^{DGH}f^{CGH}
                -\frac{1}{3} f^{BEF}f^{DEF}T^D_{ab}T^{C}_{ba}
                -2\,T^B_{ab}T^{C}_{bc}T^D_{ad}T^D_{dcdb}
                \right],
\end{align}
where $l=1/(16\pi^2)$. The fact that $\gamma_A$ is finite as $\ep\to0$ follows from the fact that it is directly related to the beta function of $g$ in the background field gauge; see \eqref{betagamma}.
On the other hand, the anomalous dimension of the scalar field is
\begin{align}
\label{eq:gammaphi}
    \gamma_\phi^{ab} &=  l \left[ 
                2\,T^A_{ac}T^A_{cb} \right]
                -l^2 \bigg[
                 \frac{3}{2}T^A_{ac}T^A_{cd} T^B_{de}T^B_{eb}
                -\frac{113}{12} T^A_{ac}T^B_{cb} f^{ACD}f^{BCD}
                +\frac{11}{12} T^A_{ac}T^B_{cb} T^A_{de}T^B_{ed}
                \nn\\&\qquad
                +\frac{1}{12} \lambda^{a cde} \lambda^{bcde}
                \bigg]
                -\frac{l^2}{\epsilon} 
                \left[
        \frac{5}{3} T^A_{ac}T^B_{cb} f^{ACD}f^{BCD}
        -\frac{1}{6} T^A_{ac}T^B_{cb} T^A_{de}T^B_{ed}
        \right],
\end{align}
with a remaining pole. Importantly, we emphasize that the anomalous dimensions of all observable quantities are necessarily finite, even when the anomalous dimensions of the fields may have remaining poles; see for instance~\cite{Manohar:2024xbh}.%
\footnote{ In the gauged scalar theory of Section~\ref{sec:scalarQCD}, the dimension-five operator proportional to the EOM of the gauged scalar field has a remaining $1/\ep$ pole at two loops, while all other anomalous dimensions do satisfy the consistency conditions.
Other cases with remaining poles in the anomalous dimensions have previously been discussed in~\cite{Jack:1990eb,Fortin:2012hn,Bednyakov:2014pia,Herren:2017uxn,Herren:2021yur,Pannell:2024sia,Jenkins:2023rtg, Jenkins:2023bls, Manohar:2024xbh,Zhang:2025ywe}.}

Finally, since we renormalize the full Green's basis, including insertions of redundant operators, we allow for further consistency checks on the counterterm calculation and basis transformations. For instance, it can explicitly be checked that the operators that can be set to zero by a field redefinition do not mix into operators in a minimal basis~\cite{Politzer:1980me}. At leading order in the Wilson coefficients, such operators are proportional to the EOM. We confirm this expected mixing structure in the specific theory of Section~\ref{sec:scalarQCD}.

\section{General Results}\label{sec:genRes}

\subsection{Notation}\label{sec:notation}

We present the complete results for the anomalous dimensions up to two loops digitally in our \texttt{GitHub} repository.%
    \footnote{
    \href{https://github.com/jasperrn/Bosonic-EFT-RGE}{https://github.com/jasperrn/Bosonic-EFT-RGE}
}
The files are divided into three folders, containing the computed counterterms in the Greens basis (including $1/\epsilon^2$ poles at two loops), the RGEs in the Green's basis (including insertions of redundant coefficients operators) and the RGEs in the minimal basis (obtained after a field redefinition), respectively.%
\footnote{In the expressions for the RGEs, we add the contributions from the wavefunction renormalization separately. This has the consequence that the resulting expression is not fully simplified using the gauge relations of Section~\ref{sec:gaugerelations}, because there might be redundancies between the contributions from the counterterms to 1PI Green's functions and the wavefunction renormalization, which were simplified separately.}
Each folder contains different files for the results at dimension four, dimension five and dimension six, where dimension six is further separated into contributions from the insertion of two dimension-five operators and from single dimension-six interactions. We provide the results in both \texttt{Form} (\texttt{.frm}) format and \texttt{Mathematica} (\texttt{.m}) format.

In the expressions, we use the notation \texttt{cPhixFyDz} and \texttt{dPhixFyDz} to denote coefficients 
$c_\mathcal{O}$ and $d_\mathcal{O}$ in \eqref{eq:lag}, where \texttt{x}, \texttt{y} and \texttt{z} are substituted by the number of scalars, field-strength tensors and derivatives in the corresponding operator, respectively. When either of these components is absent in the operator, we do not write it in the coefficient definition. To give some examples, the coefficient $d_{\phi^2FD^2}$ is denoted by \texttt{dPhi2F1D2} whereas $c_{\phi F^2}$ is written as \texttt{cPhi1F2}. 

Scalar indices are written in lowercase and gauge indices in uppercase.
Dummy indices are written \texttt{bi} and \texttt{Bi}, with integers~\texttt{i}.
The external indices (always starting with the letter $a$ or $A$) are contracted with auxiliary tensors with the appropriate symmetry properties. These tensors are either named \texttt{sym} or \texttt{asym} depending on whether it is symmetric or antisymmetric, respectively. For instance,  \texttt{sym[a1,a2] X[a1,a2]} (for some generic tensor expression \texttt{X}) prescribes the expansion \texttt{1/2*(X[a1,a2]+X[a2,a1])}. For the results of the coefficient $c^{a_1a_2a_3a_4}_{\phi^4D^2}$ this permutation tensor is \begin{align}
    &\texttt{sym[a1,a2]sym[a3,a4] X[a1,a2,a3,a4]} \leftrightarrow \nn\\
    &\texttt{1/8(X[a1,a2,a3,a4]} \texttt{+X[a1,a2,a4,a3]+X[a2,a1,a3,a4]+X[a2,a1,a4,a3]}\nn\\ 
    &\qquad\texttt{+X[a3,a4,a1,a2]+X[a3,a4,a2,a1]+X[a4,a3,a1,a2]+X[a4,a3,a2,a1])}\,.
\end{align} 
In practice, this notation has the consequence that different terms are mapped to a single term if the full expression has a particular relabeling symmetry, which significantly reduces the size of our expressions.

Before moving on to actually present some results, let us comment on the reason why, besides presenting the results in the minimal basis (obtained by setting all $d_\mathcal{O}$ to zero and performing field redefinitions, as we will describe in Section~\ref{sec:redundant}), we also give the results in in the Green's basis that includes redundant operators. While the minimal, physical results are arguably more relevant and interesting for the community, we decided to also present the Green's basis results for mainly three reasons. Firstly, we use as an internal cross-check that redundant operators only mix into other redundant degrees of freedom~\cite{Politzer:1980me}. Secondly, 
with the full Green's basis results at hand, it becomes possible to test the consistency conditions on the $1/\ep^2$ poles~\cite{tHooft:1973mfk} before performing field redefinitions, thus separating checks on the loop integrals from checks on the change of operator basis.
Thirdly, the results in the Green's basis are necessary if one wishes to project them into their own off-shell basis, to compare with off-shell calculations. 
Nevertheless, we will only present results in the minimal basis, having removed the redundant operators.

\subsection{Removing redundant operators}
\label{sec:redundant}

We perform the computations using off-shell kinematics, which means that a redundant basis of operators is necessary to capture the generated divergences. To remove the redundant degrees of freedom, one can make use of the fact that S\nobreakdash-matrix elements are invariant under field redefinitions. Therefore, 
after computing the counterterms in terms of a redundant Green's basis, one can perform the appropriate field redefinitions such that the redundant terms are reduced to those in the minimal basis. We choose the minimal basis to be the one described by the tensors labeled with a $c_\mathcal{O}$ in Eq.~\eqref{eq:lag}; the tensors labeled with a $d_\mathcal{O}$ correspond to redundant structures to be removed.  To illustrate the effect of removing a redundant operator effect, let us explicitly write the result, at dimension five, of the field redefinition that removes the redundant operator $d_{\phi^3D^2}^{abc}$:
\begin{equation}\label{eq:dim5Removal}
    c_{\phi^5,\text{ min}}^{abcde}
    = c_{\phi^5}^{abcde} - \big( 
        d_{\phi^3D^2}^{fab}\,\lambda^{fcde} + \text{ 9 permutations}
    \big) \,,
\end{equation}
where the superscript \emph{min} indicates that the coefficient relates to the operator in the minimal basis instead of the Green's basis of Eq.~\eqref{eq:lag}.
(The corresponding field redefinition also corrects the coefficients of dimension-six operators, which we do not write here.)

The necessary field redefinitions and their effect on the minimal basis coefficients have already been derived in Ref.~\cite[Sec.~3]{Fonseca:2025zjb}. Given that a user of our results does not need to perform these redefinitions -- we present the results both in the Green's and in the minimal basis -- we opted to not rewrite all the field redefinitions in this draft. Note that this was also the reason why we chose the starting Lagrangian to match that of Ref.~\cite{Fonseca:2025zjb}, including operator normalizations, such that one can directly employ their field redefinitions. 

\subsection{Partial set of general results}\label{subsec:genRes}

In this section, we present a partial set of the two-loop anomalous dimensions that we calculated in the general EFT. 
The provided expressions are meant to exemplify the general structure of the results, and to highlight two-loop features of interest in specific expressions.
For the full set of computed anomalous dimensions, we refer the interested readers to the expressions on \texttt{GitHub}.

We write the following results in the minimal basis, after field redefinitions. The coefficient $l=(16\pi^2)^{-1}$ is included to distinguish between one- and two-loop terms. Note also that we separately include the contributions from the wavefunction renormalization, $\gamma^{AB}_A$ and $\gamma^{ab}_\phi$, given in \eqref{eq:gammag} and \eqref{eq:gammaphi}, respectively. Mimicking our notation introducted in Section~\ref{sec:notation}, the results are presented contracted with a tensor $S$ or $\tilde{S}$ to indicate the symmetric (anti-symmetric) permutation symmetry of the indices.

Let us start by inspecting the running of the dimension-five coefficient $c_{\phi F^2}^{aAB}$,
\begin{align}
\label{eq:phiF2}
&S_{AB} \frac{d\,c^{aAB}_{\phi F^2}}{d\log\mu} =S_{AB} \left\{ l \left[- 4 i c^{b C A}_{\phi F^2} f^{C D B} T^{D}_{ba}\right]\right.\nonumber\\
&\left.+l^2\left[\frac{68}{3} c^{aEA}_{\phi F^2} f^{C D B} f^{CEF} f^{DGH} f^{FGH} + 2 c^{eCA}_{\phi F^2} \lambda_{acde} T^{B}_{fd} T^{C}_{fc} + 26 i c^{bGC }_{\phi F^2} f^{CD B} f^{EF A} f^{EFG} T^{D}_{ba} \right.\right.\nonumber\\
&\left.\left. \phantom{x} + 4 c^{aDC}_{\phi F^2} T^{A}_{ce} T^{B}_{de} T^{C}_{bc} T^{D}_{bd} + \frac{10}{3}i c^{b E A }_{\phi F^2} f^{CD B}  T^{C}_{ba} T^{D}_{cd} T^{E}_{cd} + \frac{2}{3} c^{aDA}_{\phi F^2} f^{CDE} f^{CEF} T^{B}_{bc} T^{F}_{bc}\right.\right.\nonumber\\
&\left.\left. \phantom{x} + 56 i c^{bFC}_{\phi F^2} f^{CD B} f^{DE A} f^{EFG}  T^{G}_{ba} - i\frac{145}{3} c^{bCA}_{\phi F^2} f^{CD B} f^{DEF} f^{EFG} T^{G}_{ba}\right]\right.\nonumber\\
&\left. \phantom{x} - \gamma^{ab}_\phi c^{bAB}_{\phi F^2} - \gamma^{BC}_A c^{aAC}_{\phi F^2} - \gamma^{AC}_A c^{aCB}_{\phi F^2} \right\}.
\end{align}
Considering a theory without kinetic mixing and restricting to one gauge group for simplicity, one can define the operator $\beta(g)/(2g) F^A_{\mu\nu}F^{A\mu\nu}$ which is not multiplicatively renormalized to all orders in perturbation theory~\cite{PhysRevD.15.1712,Nielsen:1977sy,RevModPhys.54.729}, because it is the trace of the energy-momentum tensor. 
This extends to the case when this operator is multiplied by a spectator scalar field, i.e.~a non-propagating scalar which does not have any other interactions, resulting in an operator with field content $\phi^a F^A_{\mu\nu}F^{A\mu\nu}$. Writing $c^{aAB}_{\phi F^2} = \tilde c^a_{\phi F^2} \delta^{AB}$,
it follows from this non-renormalization theorem that 
the two-loop RGE of $\tilde c^{a}_{\phi F^2}$ is given by 
\begin{equation}
    \frac{d\,\tilde c_{\phi F^2}^{a}}{d\log\mu} = 
    \left(2\beta_{g}^{(1)}+4\beta_{g}^{(2)}\right)\tilde c^{a}_{\phi F^2}\,,
\end{equation}
where the superscripts correspond to the loop order and $g$ is the gauge coupling associated with the field-strength tensor $F^A_{\mu\nu}$.
We can verify this in the general result of Eq.~\eqref{eq:phiF2} 
by keeping only the terms in which the index $a$ is on the tensor $\tilde c_{\phi F^2}$, such that the respective fields are spectators. 
At one loop, the full result is given by $-2\,\gamma_A^{(1)} \tilde c_{\phi F^2}$, as all the 1PI contributions vanish.
At two loops, the 1PI contributions do not vanish, but they are equal to $-2\,\gamma_A^{(2)}\, \tilde c_{\phi F^2}$, which combine with the wave-function renormalization to give 
$-4\,\gamma_A^{(2)} \, \tilde c_{\phi F^2}$, in agreement with the expectation.
While this identification is not immediate in the general theory -- it requires gauge relations to be applied to relate the tensor expressions -- we will observe it more transparently in Section~\ref{sec:scalarQCD}, with a specific gauge theory in mind. Finally we note that in the case of multiple gauge groups, it is the sum $\sum_A\beta(g_A)/(2g_A) F^A_{\mu\nu}F^{A\mu\nu}$ that is not renormalized~\cite{Grojean:2013kd}.

Moving to dimension-six coefficients, the running of $c_{F^3}^{ABC}$ reads
\begin{align}
\label{eq:fcubed}
&\tilde{S}_{ABC}\frac{d\,{c_{F^3}^{ABC}}}{d\log\mu} = \tilde{S}_{ABC} 
\left\{ l\left[-12 c^{BCF}_{F^3} f^{DAE} f^{DEF} - 12 c^{bDA}_{\phi F^2} c^{bEC}_{\phi F^2} f^{DBE}   \right]\right.\nonumber\\
&\left. +l^2\left[\frac{127}{3} c^{BCH}_{F^3} f^{ADE} f^{DFG} f^{EHI} f^{FGI} -18 c^{cd A E}_{\phi^2 F^2} f^{BDE} T^{C}_{cb} T^{D}_{db} + 3 c^{cd DE}_{\phi^2 F^2} f^{BCD} T^{A}_{cb} T^{E}_{db} \right.\right.\nonumber\\
&\left.\left.-\frac{29}{6} c^{BCE}_{F^3} f^{DEF} f^{DFG} T^{A}_{cb} T^{G}_{cb}+O(c_{\phi F^2} c_{\phi F^2} )\right]\right.\nonumber\\
&\left. - \gamma^{AD}_A c^{DBC}_{F^3} - \gamma^{BD}_A c^{ADC}_{F^3} - \gamma^{CD}_A c^{ABD}_{F^3}\right\},
\end{align}
where we include the contributions triggered by dimension-six coefficients at one and two loops but only insertions of dimension-five coefficients at one loop. From the non-renormalization theorem of Ref.~\cite{Cheung:2015aba}, which applies to single insertions of EFT operators, it follows that $F^3$-type operators can only be renormalized by themselves at one loop.
We confirm this result in Eq.~\eqref{eq:fcubed}, but we explicitly find that $F^3$ can be renormalized at one loop by insertions of two dimension-five operators.%
    \footnote{The non-renormalization of $F^3$ by other dimension-six operators is also predicted by the theorem of Ref.~\cite{Bern:2019wie}. The one-loop RG contribution from two dimension-five operators are consistent with the extension of this theorem to the non-linear case~\cite{Cao:2023adc}.}
We also find that $F^3$ is renormalized by other dimension-six operators at two loops.

Let us also explore a part of the one-loop renormalization of $c^{abcdef}_{\phi^6}$, namely the running triggered by the insertion of the primary operator $c^{abcd}_{\phi^4D^2}$,
\begin{align}
S_{abcdef}&\frac{d{c^{abcdef}_{\phi^6}}}{d\log\mu} \supset S_{abcdef} 
\left\{l \left[240 c^{afgh}_{\phi^4D^2} \lambda_{bchi} \lambda_{deig} - 240 c^{hfga}_{\phi^4D^2} \lambda_{bchi} \lambda_{deig} 
\right.\right.\nonumber\\
&\left.\left.
+ 2880 c^{cbhd}_{\phi^4D^2} T^{I}_{ge} T^{I}_{if} T^{J}_{ia} T^{J}_{hg}
+ 1440 c^{gehf}_{\phi^4D^2} T^{I}_{ha} T^{I}_{gi} \lambda_{bcdi} - 80 c^{afhg}_{\phi^4D^2} \lambda_{bghi} \lambda_{cdei} 
\right.\right. \nonumber\\
&\left.\left.
+ 80 c^{gfha}_{\phi^4D^2} \lambda_{bghi} \lambda_{cdei}
+480 c^{fega}_{\phi^4D^2} T^{I}_{ih} T^{I}_{gi} \lambda_{bcdh}+8640c^{gbhc}_{\phi^4D^2} T^{I}_{he} T^{I}_{ia} T^{J}_{if} T^{J}_{gd}\right]\right\}.
\end{align}
In Ref.~\cite{Cao:2021cdt}, it was shown that this mixing vanishes in the $\mathcal{O}(n)$ scalar model. Indeed, we recover this result by the projection $\lambda_{abcd}\rightarrow(\delta_{ab}\delta_{cd}+\delta_{ac}\delta_{bd}+\delta_{ad}\delta_{bc})$ and setting $T^A_{ab}=0$. This zero can be used as a consistency check on the applied field redefinition.
However, our general results show that this non-renormalization result does not extend to different scalar theories.

\subsection{Renormalization from coefficients with positive mass dimension}\label{sec:posMass}

Our starting Lagrangian in Eq.~\eqref{eq:lag} does not include masses or $\phi^3$ operators, so our general results do not include the mixing of higher-dimensional operators into lower-dimensional ones. 
Remarkably, such contributions can be derived from our general expressions using the dummy field method~\cite{Martin:1993zk,Luo:2002ti,Schienbein:2018fsw,Bednyakov:2021ojn}. In particular, with appropriate normalizations, the Feynman rule of a $\phi^k$ operator in the general theory is simply $c^{a_1\cdots a_k}$. For this reason, diagrams with an insertion of $\phi^k$ can be related to diagrams with an insertion of $\phi^{k-n}$, by taking $n$ indices to correspond to spectator fields without interactions. 
This can also be phrased as a shift on the external fields, $\phi^a \rightarrow \phi^a+x^a$, where $x^a$ is a non-propagating field. 

To exemplify the dummy field method, we start with the renormalization of $\phi^6$ into itself proportional to $g^2\lambda^{abcd}$,
\begin{align}
\label{eq:c6result}
S_{abcdef}&\frac{d\,{c_{\phi^6}^{abcdef}}}{d\log\mu} \supset S_{abcdef}\left\{l^2\left[ 
-120 \, c_{\phi^6}^{cdegij} \lambda^{bghi} T^I_{hf}T^I_{ja} 
- 60 \, c_{\phi^6}^{adefij} \lambda^{bcgh} T^I_{ig}T^I_{jh}\right. \right.\nonumber\\
& \left.\left.
+60 \, c_{\phi^6}^{adefhj} \lambda^{bcgh} T^I_{ig}T^I_{ji}  
+ 120 \, c_{\phi^6}^{defhij} \lambda^{bcgh} T^I_{ia}T^I_{jg} 
- 60 \, c_{\phi^6}^{cdefij} \lambda^{bghi} T^I_{jh}T^I_{ga}  \right]\right\}.
\end{align}
To recover the two-loop contribution to the RGE of $\lambda^{abcd}$ from $c_{\phi^6}^{abcdef}$ proportional to $g^2$,
we contract two of the open indices in Eq.~\eqref{eq:c6result} with dummy fields, $x^e x^f$. Note that to perform this contraction one should first expand the right-hand side to include all permutations of the external indices encoded in $S_{abcdef}$. 
Using the shorthand $c_{\phi^6}^{abcdxx} = 
c_{\phi^6}^{abcdef} x^e x^f$, etc., we identify
$c_{\phi^6}^{abcdxx}= -2 \lambda^{abcd}$ (the minus sign arises from the relative sign in the Lagrangian) and $\lambda^{abxx} = 2\, m_{ab}^2$, for a mass term defined by
$\mathcal{L}\supset -m^2_{ab}/2 \,\phi^a\phi^b$. We set $T^A_{ax} = 0$, and we consider only terms proportional to $m^2$, neglecting the terms proportional to $c_{\phi^5} c_{\phi^3}$ or $\lambda^2$, which can also be obtained in this way.
This results in
\begin{align}
\label{eq:c6tolambdaresult}
S_{abcd}\frac{d\,{c_{\phi^6}^{abcdxx}}}{d\log\mu}=S_{abcd}\left(
-2\,\frac{d\,{\lambda^{abcd}}}{d\log\mu}\right)
&\supset \frac{(2)(4!)}{6!} S_{abcd} \left\{l^2\left[
- 120 c_{\phi^6}^{adbcij} m^2_{gh} T^I_{ig}T^I_{jh} \right.\right.
\nn\\&\left.\left.
+ 120 c_{\phi^6}^{adbchj} m^2_{gh} T^I_{ig}T^I_{ji}
+ 240 c_{\phi^6}^{dbchij} m^2_{gh} T^I_{ia}T^I_{jg}  \right]\right\}.
\end{align}
We use this method to reproduce the RG mixing contributions proportional to the Higgs mass in the SMEFT in Section~\ref{sec:SMEFT}. 

\section{Projecting onto specific theories}\label{sec:proj}
In this section, we exemplify the extraction of results for specific theories from our general expressions. In particular, we reproduce the results of Ref.~\cite{Born:2024mgz} in  the bosonic SMEFT, which is an important cross-check on our calculation. Since this calculation does not involve dimension-five operators, we also consider an $SO(N_c)$ scalar gauge theory in Section~\ref{sec:scalarQCD}, for which we perform a series of internal consistency checks.

\subsection{SMEFT}\label{sec:SMEFT}
The SMEFT is arguably one of the most interesting phenomenological effective theories onto which to project our results. Ref.~\cite{Born:2024mgz} produced the two-loop RGEs for the bosonic sector of the SMEFT using functional methods.
Validating these results by projecting our general expressions to the SMEFT constitutes a powerful cross-check we performed the calculation in a different manner. We will not write the SMEFT RGEs at two loops here, as they can be accessed in Ref.~\cite{Born:2024mgz}. However, we comment on the procedure to project the general results onto the SMEFT, which can be adapted to project into other specific theories; see also~\cite[Sec.~5]{Aebischer:2025zxg} for a discussion on the projection of the general theory to the SMEFT or other theories.

The SMEFT respects the gauge group $SU(3)\times SU(2) \times U(1)$  and as such the structure constants $f^{ABC}$ read
\begin{align}
    &&f^{111} &= 0\,, & A^1_\mu &= B_\mu\,, \nn\\
    \text{ for } A,B,C\in \{2,3,4\}, \  &&f^{ABC} &= g_2 \, \epsilon^{IJK}\,, & A^A_\mu &= W_\mu^{I}\,,
    \nn\\ \label{eqG}
    \text{ for } A,B,C\in \{5,...12\},  &&f^{ABC} &= g_s\, f_s^{\tilde A \tilde B \tilde C}\,,
        &A^A_\mu &= G_\mu^{\tilde A}\,,
\end{align}
where we drop the tildes on the uppercase $SU(3)$ indices in what follows.
Here, $\epsilon^{IJK}$ is the Levi-Civita tensor and $f_s$ are the $SU(3)$ structure constants. Moreover, the structure constants vanish whenever indices are mixed between different groups. The Higgs doublet can be written in terms of its components as
\begin{equation}\label{phitoH}
    H = \frac{1}{\sqrt{2}} \pmx{\phi^1 + i \phi^3\\\phi^2+i\phi^4},
\end{equation}
where the real scalar fields $\phi^a$ are identified with the  components of the general scalar multiplet in the general EFT.

With this identification, the projection of the remaining tensors can be computed by expanding the target theory in its components and matching to the general tensors. Besides the aforementioned Ref.~\cite{Aebischer:2025zxg}, geometric approaches to EFTs also treat the Higgs as a vector of real scalar fields~\cite{Alonso:2015fsp,Helset:2022pde}.

For the SM, the generators acting on the four real components of the Higgs can be projected, following Ref.~\cite{Aebischer:2025zxg}, as
\begin{align}
\label{eq:gaugeprojectionsmeft}
T^1 = \frac{i g_1}{2}\begin{pmatrix}0&\mathbbm{1}\\-\mathbbm{1}&0
    \end{pmatrix},\qquad T^{(I+1)}=\frac{i g_2}{2}\begin{pmatrix}\mathrm{Im}(\sigma^I)&\mathrm{Re}(\sigma^I)\\-\mathrm{Re}(\sigma^I)&\mathrm{Im}(\sigma^I)
\end{pmatrix},
\end{align}
where $\sigma^I$ are the Pauli matrices, with $I\in\{1,2,3\}$.
The remaining interactions in the bosonic sector of the SMEFT  (bSMEFT) that we want to project onto include the renormalizable quartic interaction and dimension-six terms:
\begin{align}
&\mathcal{L}_\text{bSMEFT}=   -\frac14 B_{\mu\nu} B^{\mu\nu}  -\frac14 W^A_{\mu\nu} W^{A\mu\nu} -  \frac14 G^A_{\mu\nu} G^{A\mu\nu} + D_\mu H^\dagger D_\mu H -\lambda (H^\dagger H)^2\,\nonumber\\ 
&+ c_H (H^\dagger H)^3 + c_{H\square} (H^\dagger H)D^2(H^\dagger H)+c_{HD} (H^\dagger D_\mu H)(D_\mu H^\dagger H)  \nonumber\\
& + c_{HB} (H^\dagger H) B_{\mu\nu}  B^{\mu\nu}+c_{HW} (H^\dagger H) W^A_{\mu\nu} W^{A\mu\nu} +c_{HWB} (H^\dagger \sigma^A H) W^A_{\mu\nu} B^{\mu\nu} \nonumber \\
&+c_{HG} (H^\dagger H) G^A_{\mu\nu} G^{A\mu\nu} + c_G f_s^{ABC}G^{A\nu}_{\mu}G^{B\rho}_{\nu} G^{C\mu}_{\rho}+ c_W \epsilon^{ABC}W^{A\nu}_{\mu}W^{B\rho}_{\nu} W^{C\mu}_{\rho}\,.
\end{align}

The relations to map the tensors of the general EFT to the bSMEFT coefficients read
\begin{align}
\label{eq:smeftprojections}
&\lambda^{abcd}\rightarrow 2\lambda (\delta_{ab}\delta_{cd}+\delta_{ac}\delta_{bd}+\delta_{ad}\delta_{bc})\,,
\hspace{1cm}
c^{IJK}_{F^3} \rightarrow 6 c_{W^3} \epsilon^{IJK} \,,\nonumber\\[2mm]
&c^{ABC}_{F^3} \rightarrow 6 c_{G^3} f^{ABC}\,,
\hspace{3.74cm}
    c^{abIJ}_{\phi^2 F^2} \rightarrow
        2\, c_{W^2H^2}\delta^{IJ}\delta^{ab}        
        \,,\nonumber\\[2mm]
        &   c^{ab11}_{\phi^2 F^2} \rightarrow 2 \, c_{B^2H^2} \delta^{ab}\,,
        \hspace{3.73cm}
        c^{abI1}_{\phi^2 F^2} \rightarrow c_{WBH^2}X^{I}_{ab}\,, \nonumber\\[2mm]
       & c_{\phi^4 D^2}^{abcd} \rightarrow 
    \left(\frac{c_{H\square}}{3}
    -\frac{c_{HD}}{12} \right)
    \left(\delta^{ab}\delta^{cd}
    -\frac12 \delta^{ac}\delta^{bd}
    -\frac12 \delta^{ad}\delta^{bc} \right)
    +
    \left(
    \frac{c_{HD}}{8} \right)
    \left( \Omega^{ad}\Omega^{bc} - \Omega^{ac}\Omega^{db}
    \right)\,,\nonumber\\
    &c_{\phi^6}^{abcdef} \rightarrow (6 c_{H} - 16 c_{H\square}\lambda + 4 c_{HD}\lambda)(\delta_{ab}\delta_{cd}\delta_{ef}+\text{14 permutations})\,,
\end{align}
where lowercase indices go from 1 to 4, $I,J,K\in\{2,3,4\}$ and $A,B,C\in\{5,...12\}$ and we define the matrices
\begin{align}
    &X^2 =  \begin{pmatrix} \sigma^1 & 0\\0&\sigma^1 \end{pmatrix},
    &&X^3 = \begin{pmatrix} 0 & i\sigma^2\\-i\sigma^2&0 \end{pmatrix},
    &&X^4 =  \begin{pmatrix} \sigma^3 & 0\\0&\sigma^3 \end{pmatrix}\,,
    && \Omega = 
    \begin{pmatrix}
        0&\mathbbm{1}_{2\times2}\\
        -\mathbbm{1}_{2\times2}&0
    \end{pmatrix}.
\end{align}
With these rules, we projected the general results onto the bSMEFT and reproduced the one- and two-loop results from Ref.~\cite{Born:2024mgz}.

\paragraph{Including massive parameters.}
While we will not explicitly transcribe the RGEs of the bSMEFT at two loops, let us take the opportunity to showcase another application of our general results which we described in Section~\ref{sec:posMass}. Although we did not include terms with positive mass dimension in our general EFT, we can recover terms proportional to the Higgs mass from our results. 
In Section~\ref{sec:posMass}, we explained how to recover the renormalization of $\lambda$ proportional to $c_{\phi^6}$. Indeed, projecting Eq.~\eqref{eq:c6result}
onto the SMEFT results in
\begin{equation}
\frac{d\,c_H}{d\log\mu}\supset l^2 (144 g_1^2+432 g_2^2) \, \lambda \, c_H\,,
\end{equation}
which agrees with Ref.~\cite{Born:2024mgz} (note the factor of two difference between their conventions and ours). 
To recover the running of $\lambda$ triggered by $c_H$ we need to shift the external indices in $\lambda^{abcd}$ in Eq.~\eqref{eq:c6result} which results in Eq.~\eqref{eq:c6tolambdaresult}; projecting the latter onto the SMEFT, taking $m^2_{ab}\rightarrow -\mu^2_H \delta_{ab}$, with $\mathcal{L}\supset \mu_H^2 H^\dagger H$, results in
\begin{equation}
\frac{d\,\lambda}{d\log\mu}\supset l^2 (48 g_1^2+144 g_2^2) \,\mu_H^2 \, c_H \,,
\end{equation}
which once again agrees with the results of Ref.~\cite{Born:2024mgz}.
The results in the general theory thus efficiently encode contributions from massive parameters, even without explicitly including these in the performed calculation.

\subsection{$SO(N_c)$ scalar QCD coupled to $N_f$ scalar singlets}\label{sec:scalarQCD}

To further illustrate the determination of anomalous dimensions in specific theories from our general expressions, we now consider the reduction to a gauged scalar theory with $SO(N_c)$ gauge symmetry. 
We include one vector of scalar fields in the fundamental representation, $\phi^\alpha$ with $\alpha=1,...,N_c$, and $N_f$ singlet scalars with arbitrary global symmetry (i.e.~without any flavor symmetry imposed), $\varphi^i$ with $i=1,...,N_f$. 

The calculation of this section also serves to cross-check our results in two ways, as described in Section~\ref{sec:cross}. 
Firstly, besides extracting the anomalous dimensions directly from the results in the general theory, we recompute the anomalous dimensions from the counterterms in the general theory. This confirms the consistency between the one-loop $1/\ep$ pole and the two-loop $1/\ep^2$ pole in these counterterms. (As we will see below, not all consistency conditions are satisfied, because we did not include the RG running of the gauge-fixing parameter $\xi$.) 
Secondly, by extracting the mixing matrix in the full Green's basis, we expose the expected mixing structure: operators that are proportional to the EOM do not mix into operators in the minimal basis.

\paragraph{Dimension four.}
The relevant Lagrangian at dimension four is 
\begin{align}
    \mathcal{L}^{(4)} &= 
    -\frac14 F^A_{\mu\nu} F^{A\mu\nu} 
    + \frac12 D_\mu \phi^\alpha D^\mu \phi^\alpha 
    + \frac12 \partial_\mu \varphi^i \partial^\mu \varphi^i 
    \nonumber\\&\quad
    -\frac{\lambda_1}{8} (\phi^\alpha \phi^\alpha)^2
    -\frac{\lambda_2^{ij}}{4} (\phi^\alpha \phi^\alpha)
            \varphi^i \varphi^j
    -\frac{\lambda^{ijkl}_3}{4!}\varphi^i 
    \varphi^j \varphi^k \varphi^l\,.
\end{align}
This theory can be related to a Lagrangian with a single scalar multiplet $\Phi$ by identifying $\vec \Phi = (\vec \phi,\,\vec \varphi)$. In terms of indices, we write $\Phi^\alpha = \phi^\alpha$ for $\alpha=1,...,N_c$ and $\Phi^{i+N_c} = \varphi^i$ for $i=1 ,...,N_f$.
In addition, we restrict the gauge index $\scriptstyle A$ to run between $1$ and 
$\frac{1}{2}N_c(N_c-1)$.
The relation to the general theory~\eqref{eq:lag} is then given by the replacement rules 
\begin{align}
    T^A_{\alpha\beta} &= g\,T^A_{\alpha\beta}\,, 
    \quad 
    T^A_{\alpha i} = 0\,, 
    \quad 
    T^A_{ij} = 0\,, \nn\\
    \lambda^{\alpha\beta\gamma\delta} &= \lambda_1\,(\delta^{\alpha\beta} \delta^{\gamma\delta} + 
    \delta^{\alpha\gamma}\delta^{\beta\delta} + \delta^{\alpha\delta}\delta^{\beta\gamma})\,,\nn\\
    \lambda^{\alpha\beta ij} &= \lambda_2^{ij}\delta^{\alpha\beta}
    \,,\nn\\
    \lambda^{ijkl} &= \lambda_3^{ijkl}\,.
\end{align}
The two-loop RGEs at dimension four can then be extracted to be
{\allowdisplaybreaks
    \begin{align}
    \gamma_\phi & = 
        l\left[\frac{1}{2}(N_c-1)g^2 \right]
        -\frac{l^2}{4}\left[ 
            \lambda_1^2(N_c+2)
            -\frac{1}{6}g^4(N_c-1) (26 N_c -57)
            +\lambda_2^{ij}\lambda_2^{ij} \right]\nn\\&\quad
        -\frac{l^2}{48\epsilon}\Big[ 
            (N_c-1)(10N_c-21)g^4
        \Big], \nn\\[3mm]
    \gamma_\varphi^{ij} &= -\frac{l^2}{4}\left[ 
    \lambda_2^{ik}\lambda_2^{jk}N_c
    +\frac{1}{3}\lambda_3^{iklm}\lambda_3^{jklm}
    \right]\,,\nn\\[3mm]
    \gamma_A &= l\,g^2 \left[ 
         \frac{11}{6} N_c
         -\frac{15}{4}
    \right]
    + \frac{l^2\,g^4}{12} \Big[
       141-140N_c + 34 N_c^2
    \Big] \,,\nn\\[3mm]
    \beta_g &= - g \ep - g\,\gamma_A
    \,,
    \nn\\[3mm]
\beta_{\lambda_1}
&= 
        -2 \epsilon \lambda _1 + 
l \left[\frac{3}{4} g^4 \left(N_c-1\right)-3 g^2 \lambda _1
   \left(N_c-1\right)+\lambda _1^2 \left(N_c+8\right)+(\lambda _{2}^{ij})^2\right]
   +O(l^2)
 \,,
\nn\\[3mm]
\beta_{\lambda_2}^{ij} &= -2\ep \lambda_2^{ij}
+l \left[-\frac{3}{2} g^2 \left(N_c-1\right) \lambda
   _{2}^{ij}+\lambda_1 \left(N_c+2\right) \lambda _{2}^{ij}+4
   \lambda _{2}^{ik} \lambda _{2}^{jk}+\lambda _{2}^{kl}
  \lambda_3^{ijkl} \right]
  +O(l^2)\,,
   \nn\\[3mm]
\beta_{\lambda_3}^{ijkl} &= 
-2\ep \lambda_{3}^{ijkl} + 
l \left[N_c \lambda _{2}^{ij} \lambda _{2}^{kl}
   +\lambda
   _{3}^{ijmn} \lambda _{3}^{klmn}
   + \text{ 2 perms}
   \right]
   +O(l^2)\,.
 \label{betaScalarQCD}
    \end{align}
We provide the two-loop expressions for the beta functions of $\lambda_i$ in Appendix~\ref{app:RGE}.
Three-loop anomalous dimensions and beta functions can be extracted from the general results at dimension four of Ref.~\cite{Steudtner:2024teg}, while six-loop results can be extracted from the scalar theory results of \cite{Bednyakov:2021ojn} in the limit $N_c=0$ and $g=0$.

We note that $\gamma_\phi$ has a remaining $1/\ep$ pole at two loops. This is because the anomalous dimension of the field depends on the gauge-fixing parameter $\xi$, but we have ignored this effect.

\paragraph{Dimension five.}
The Lagrangian at dimension five is
\begin{align}
    \mathcal{L}^{(5)} &= 
    c^i_{\varphi F^2} \, \varphi^iF^A_{\mu\nu}F^{A\mu\nu}
    +
    \frac{d^i_{\varphi \phi^2,1}}{2} \,
        \partial^2\varphi^i \phi^\alpha \phi^\alpha 
    +
    d^i_{\varphi \phi^2,2}\,
        \varphi^i (D^2\phi^\alpha) \phi^\alpha 
    +
    \frac{d^{ijk}_{\varphi^3}}{6} \partial^2\varphi^i \varphi^j \varphi^k \nn\\&\quad 
    +
    \frac{c^i_{\varphi \phi^4}}{8}
    \varphi^i (\phi^\alpha \phi^\alpha)^2
    +
    \frac{c^{ijk}_{\varphi^3 \phi^2}}{12}
    \varphi^i \varphi^j \varphi^k(\phi^\alpha \phi^\alpha)
    +
    \frac{c^{ijklm}_{\varphi^5}}{120}
    \varphi^i \varphi^j \varphi^k \varphi^l \varphi^m\,.
    \label{dim5SpecificGreens}
\end{align}
These operators can be obtained from the general theory \eqref{eq:lag} using
\begin{align}
    c_{\phi F^2}^{iAB} &= c_{\varphi F^2}^i \delta^{AB}\,,
     & d_{\phi^3D^2}^{i \alpha \beta} &= d^i_{\varphi\phi^2,1}\delta^{\alpha\beta}\,,
    \nn\\
    d_{\phi^3D^2}^{\alpha \beta i} 
    &= d_{\phi^3D^2}^{\alpha i \beta}= d^i_{\varphi\phi^2,2} \delta^{\alpha\beta}\,,
&
     d_{\phi^3D^2}^{ijk} &= d_{\varphi^3}^{ijk}\,,
         \nn\\
     c_{\phi^5}^{\alpha \beta \gamma \delta i} &=
     c_{\varphi\phi^4}^i
     \left( \delta^{\alpha\beta}\delta^{\gamma\delta}
     +\delta^{\alpha\gamma}\delta^{\beta\delta}
     +\delta^{\alpha\delta}\delta^{\beta\gamma}
     \right)
     ,&
    c_{\phi^5}^{\alpha \beta ijk} &= c_{\varphi^3\phi^2}^{ijk}\delta^{\alpha\beta} \,,\nn\\
    c_{\phi^5}^{ijklm} &= c_{\varphi^5}^{ijklm}\,, \label{eq:reduceSpecific}
\end{align}
while all coupling constant tensors with an odd number of Greek indices are zero.
However, instead of using the Green's basis~\eqref{dim5SpecificGreens}, we will present the results in a basis in which the redundant operators multiply EOM operators, 
\begin{align}
    \lambda^{(5)} \supset &
    \frac{d^i_{\varphi\phi^2,1}}{2} (\phi^\alpha \phi^\alpha) \, \text{EOM}_\varphi^i
    + d^i_{\varphi\phi^2,2} \varphi^i \phi^\alpha \, \text{EOM}_\phi^\alpha 
    + d^{ijk}_{\varphi^3} \varphi^j \varphi^k \, \text{EOM}_\varphi^i\,,
\end{align}
with 
\begin{align}
    \text{EOM}_\varphi^i &= \partial^2\varphi^i + \frac{\lambda_2^{ij}}{2}\varphi^j(\phi^\alpha \phi^\alpha) + \frac{\lambda^{ijkl}}{6} \varphi^j \varphi^k \varphi^l\,,\nn\\
    \text{EOM}_\phi^\alpha &= D^2 \phi^\alpha 
        +\frac{\lambda_1}{2} \phi^\alpha (\phi^\beta \phi^\beta) 
        +\frac{\lambda_2^{ij}}{2} \phi^\alpha \varphi^i \varphi^j 
        \,.
        \label{eq:EOMs}
\end{align}
The anomalous dimensions in this basis can be obtained from the general theory results, after first mapping them to the Green's basis using \eqref{eq:dim5Removal}, followed by the replacement rule~\eqref{eq:reduceSpecific}. Equivalently, one can first apply~\eqref{eq:reduceSpecific}, and use the coupling redefinitions
\begin{align}
    c^{i}_{\varphi\phi^4,\,\text{min}} &= 
    c^{i}_{\varphi\phi^4} - 2 \, d_{\varphi\phi^2,1}^j \lambda_2^{ij}
            - 4 \, d_{\varphi\phi^2,2}^{i} \lambda_1 \,,
    \nn\\
    c^{ijk}_{\varphi^3\phi^2,\,\text{min}} &= 
    c^{ijk}_{\varphi^3\phi^2} 
        - \left( d^{lij}_{\varphi^3} \lambda_2^{lk} + \text{ 2 perms} \right)
            - 2 \left( d_{\varphi\phi^2,2}^{i} \lambda_2^{jk} + \text{ 2 perms} \right)
    - d_{\varphi\phi^2,1}^l \lambda_3^{lijk}\,,
    \nn\\
    c^{ijklm}_{\varphi^5,\,\text{min}} &= 
    c^{ijklm}_{\varphi^5} - 
        \left(d^{nij}_{\varphi^3}\lambda_{nklm} + \text{ 9 perms}\right)
        \,.
\end{align}
In the resulting basis, the operators parametrized by $c_{\mathcal{O},\text{\,min}}$ tensors correspond to the operators in a minimal basis, because the operators proportional to the EOM can be removed by a field redefinition.

For brevity, we present the resulting anomalous dimensions after imposing $O(N_f)$ flavor symmetry at dimension four, using
\begin{align}
    \lambda_3^{ijkl} = \frac{\lambda_3}{3}\left(\delta^{ij}\delta^{kl}
        +\delta^{ik}\delta^{jl}
        +\delta^{il}\delta^{jk}\right),
        &&
    \lambda_2^{ij} = \lambda_2\,\delta^{ij}\,,
\end{align}
but we keep the flavor structure of the dimension-five couplings general. In this way, operators in multiple representations of $O(n)$ are captured at once. The results for the operators in the minimal basis are
{\allowdisplaybreaks
\begin{align}
    \frac{d\,c^i_{\varphi F^2}}{d\log\mu} &= 
    -\ep c^i_{\varphi F^2}
    -l\left[\frac{g^2}{6}(22N_c-45)c^i_{\varphi F^2}\right] 
    \nn\\&\hspace{-10mm}
    +l^2\left[
   c^i_{\varphi F^2} 
   \left(\frac{1}{36}\left(9 \lambda
   _2^2 N_c+\lambda _3^2
   \left(N_f+2\right)\right)-
   g^4
   \left(\frac{34 N_c^2}{3}-\frac{140
   N_c}{3}+47\right)-g^2 \lambda
   _2\right)\right],
   \nn\\[3mm]
 \frac{d\,c^i_{\varphi \phi^4}}{d\log\mu} &= 
    -3\ep c^i_{\varphi \phi^4}
    +l\bigg[2 \lambda _2 c_{\varphi^3\phi^2}^{ijj}+c_{\varphi\phi^4}^i \left(3g^2 \left(1- N_c\right)+2 \lambda _1 \left(N_c+8\right)+8 \lambda _2\right)
    \nn\\&\nn\hspace{4cm}
    +c_{\varphi F^2}^i \left(3g^4 \left(1- N_c\right)+6 g^2 \lambda _1 \left(N_c-1\right)\right)\bigg] + O(l^2)\,,
    \nn\\[3mm]
\frac{d\,c^{ijk}_{\varphi^3 \phi^2}}{d\log\mu} &= 
    -3\ep c^{ijk}_{\varphi^3 \phi^2}
 +l\bigg[
\left(
 -\frac{3}{2} \left(N_c
 -1\right) g^2
 +\left(N_c
 +2\right) \lambda _1
 +2 \left(6 \lambda _2
 +\lambda _3\right)\right) c_{\varphi^3\phi^2}^{ijk}
 +\lambda _2 c_{\varphi^5}^{ijkll}
 \nn\\&\hspace{-10mm}
 +\left(3 c_{\varphi F^2}^i \left(N_c
    -1\right) \lambda _2 \delta _{jk} g^2
 +c_{\varphi\phi^4}^i \left(N_c
    +2\right) \lambda _2 \delta _{jk}
 +\frac{1}{3} \lambda _3 \delta _{jk} c_{\varphi^3\phi^2}^{ill} 
+\text{ 2 perms}
 \right)
  \bigg] +O(l^2)\,,
 \nn\\[3mm]
 \frac{d\,c^{ijklm}_{\varphi^5}}{d\log\mu} &= 
    -3\ep c^{ijklm}_{\varphi^5}
    +l\bigg[\frac{20}{3}\lambda_3 c_{\varphi^5}^{ijklm}
    +\bigg( c_{\varphi^3\phi^2}^{ijk} \delta^{lm} \lambda_2N_c
            +\frac{1}{3} c_{\varphi^5}^{ijknn}\delta^{lm}
            + \text{ 9 perms} \bigg)\bigg]
 \nn\\&\hspace{-14mm}
+l^2\bigg[-\frac{5}{36}c_{\varphi^5}^{ijklm} \left(63 \lambda _2^2 N_c
 +\lambda _3^2 \left(7 N_f
 +142\right)\right)
\nn\\&\hspace{-14mm}
    +\bigg( 
    c_{\varphi^3\phi^2}^{ijk} \delta _{lm} \left(2 g^2 \lambda _2 \left(N_c
 -1\right) N_c 
 -\frac{1}{2} \lambda _2 \left(16 \lambda _2
 +5 \lambda _3\right) N_c \right)
 -\frac{31}{18} \lambda _3^2 \delta _{lm} c_{\varphi^5}^{ijknn}
 + \text{ 9 perms}\bigg)\bigg]\,,
 \label{eq:dim5ScalarQCD}
\end{align}
where we provide some relatively large two-loop expressions in Appendix~\ref{app:RGE} in Eq.~\eqref{eq:dim5ScalarQCDTwoLoop}.
} 
A few comments on these anomalous dimensions are in order.
\begin{itemize}
    \item We emphasize that these anomalous dimensions have been computed in the full Green's basis, including operators that can be removed by field redefinitions. 
    We use the absence of $d_\mathcal{O}$-dependence in the above expressions in the minimal basis as a consistency check on the calculation.

    \item We note that the anomalous dimension of $c_{\varphi F^2}^i$ is given by $-2\gamma_A^{(1)} -4\gamma_A^{(2)}$ in the limit that $\varphi$ is a spectator field (i.e.~$\lam_2 = \lam_3=0$), where $\gamma_A^{(1)}$ and $\gamma_A^{(2)}$ are the one- and two-loop anomalous dimensions of the gauge field, respectively. This follows from the fact that $\beta_g/(2g) \varphi F^2$ is not renormalized in this limit, as discussed in Section~\ref{subsec:genRes}.

    \item Just like the anomalous dimension of the gauge field, $\gamma_\phi$, the anomalous dimension of $d_{\varphi\phi^2,2}^{i}$ has a remaining $1/\ep$ pole, 
    thus violating the consistency conditions discussed in Section~\ref{sec:cross}.
    The corresponding operator is proportional to the EOM of the gauge field, EOM$_\phi$; defined in Eq.~\eqref{eq:EOMs}. The explicit expression takes the form
    \begin{equation}
    \frac{d\,d^i_{\varphi \phi^2,2}}{d\log\mu} = l^2\left[\frac{3 g^4 \left(N_c^2-3 N_c+2\right) c_{\varphi F^2}^i}{8 \epsilon } \right]+ O(\ep^0)\,.
    \end{equation}
    The failure of the $1/\ep$ poles to cancel means that the one-loop anomalous dimension of $d_{\varphi\phi^2,2}^{i}$ depends on the gauge-fixing parameter $\xi$. We have set $\xi=1$ throughout the calculation, but it would be necessary to keep $\xi$ general and consider its renormalization to obtain finite expressions. We emphasize that none of the parameters in the minimal basis ($c_{\mathcal{O},\text{\,min}}$) are sensitive to $\xi$, because these parameters correspond to observable quantities.
    Indeed, we confirm that the consistency conditions are satisfied for their RGEs.
\end{itemize}

\section{Outlook}\label{outlook}

We have developed a general framework for the multiloop renormalization of general bosonic EFTs, including the identification of an operator basis, the extraction of UV divergences using the R$^*$ method, and the simplification of the results using tensor relations imposed by gauge symmetry. Our computed general template expressions for the two-loop anomalous dimensions of operators up to dimension six can be specialized to any theory of scalar particles and gauge bosons in $4-2\ep$ dimensions. Such an encoding of the results is efficient, not only because it includes results for all operators in a large array of theories, but also because it allows the derivation of contributions from parameters with positive mass dimension, without including them explicitly in the calculation. 
As an important application of our results, by imposing the particle content and gauge symmetry of the SMEFT without fermions, we reproduced the recently-computed RGEs from Ref~\cite{Born:2024mgz}.

This work opens various avenues for future research. Natural extensions of our calculation would be to consider higher loops or higher mass dimensions, and the inclusion of fermionic particles. 
For instance, completing the one-loop SMEFT RGEs at dimension eight would be useful for high-energy phenomenology because there exist observables that receive their leading corrections at dimension eight. 
In addition, the general RGEs at dimension eight would allow to further shed light on the interplay between positivity bounds on Wilson coefficients and their renormalization group evolution~\cite{Chala:2021wpj,Chala:2023jyx,Chala:2023xjy,Liao:2025npz}.
Moreover, performing the renormalization with general field indices allows to make the connection with geometric approaches to field theory~\cite{Helset:2022pde,Jenkins:2023rtg,Jenkins:2023bls,Assi:2025fsm}, which would be worth exploring.

Higher-loop results would be desirable for extractions of the scaling dimensions of composite operators in conformal field theories, extending the pure scalar results of Refs.~\cite{Henriksson:2025hwi,Henriksson:2025vyi} to theories that include gauge bosons.
For example, Refs.~\cite{Ihrig:2019kfv,Chala:2025crd} recently studied the Abelian Higgs Model in 3d using the $\ep$ expansion.
By virtue of the generality of template expressions, for a specific global symmetry, our results encode the scaling dimensions of operators in \emph{all} representations of that group.
In such studies, an exhaustive determination of the operator spectrum would require the additional inclusion of operators with non-zero Lorentz spin in the calculation.
Precision perturbative results then allow for theoretical studies such as the comparison between perturbative and non-perturbative results (see e.g.~\cite{Henriksson:2022gpa,Henriksson:2025hwi}), and for justifying assumptions in conformal bootstrap studies, as in Ref.~\cite{Kousvos:2025ext}.

It would be also be useful to extend the program of the renormalization of general EFTs to $3-\ep$ dimensions, with recent progress made on the beta functions of scalar-fermion theories in~\cite{Schroder:2025rka} (see also~\cite{Jack:2016utw,Fraser-Taliente:2024rql}).
Applications would include, for example, the study of thermal field theories through 3d EFTs, particularly relevant for the accurate study of phase transitions~\cite{Ekstedt:2024etx,Chala:2024xll,Chala:2025aiz,Chala:2025cya},
as well as the $\ep=3-d$ expansion of scaling dimensions in conformal field theory~\cite{STEPHEN197389,PhysRevB.18.5099,PhysRevE.60.2071,Hager:2002uq,Basu:2015gpa,Badel:2019khk,BenAliZinati:2021rqc,Kapoor:2021lrr}, enabling comparisons with the conformal bootstrap~\cite{Henriksson:2025kws}.

In closing, we hope that our work, together with the abundance of new results emerging from the community effort in general theories, will 
lay the foundation for future phenomenological studies and uncover deeper insights into the structure of EFTs.

\section*{Acknowledgements}

We are grateful for encouragement from F.~Herzog and for access to his implementation of the R$^*$ method.
JRN thanks J.~Henriksson, F.~Herzog and S.~R.~Kousvos for collaboration on related topics.
JRN is supported by the Yushan Young Scholarship 112V1039 from the Ministry of Education (MOE) of Taiwan, by the National Science and Technology Council (NSTC) grant 113-2112-M-002-038-MY3, and by the NSTC grant 114-2923-M-002-011-MY5. This work has been partially funded by the Eric \& Wendy Schmidt Fund for Strategic Innovation through the CERN Next Generation Triggers project under grant agreement number SIF-2023-004.

\appendix

\section{Renormalization group equations}\label{app:RGE}

In this appendix, we list some of the anomalous dimensions of the $SO(N_c)$ scalar gauge theory that were too large to include in the main text.
The two-loop beta functions in Eq.~\eqref{betaScalarQCD} are given by
\begin{align}
&
\beta_{\lambda_1} \Big|_{l^2} = 
l^2 \bigg[
\frac{1}{6} g^6 \left(19 N_c^2-70 N_c+51\right)
 -\frac{1}{12} g^4 \lambda _ 1 \left(64 N_c^2-271 N_c+207\right)
 \nn\\&\qquad\qquad
 +2 g^2 \lambda _ 1^2 \left(N_c^2+4 N_c-5\right)
 -4 \lambda_2^{ij}\lambda_2^{ik}\lambda_2^{jk}
 -5 \lambda _ 1 (\lambda_2^{ij})^2
 -3 \lambda_1^3 \left(3 N_c+14\right)
 \bigg]\,,
\end{align}
\begin{equation}\label{eq:betalam2L2}
\resizebox{\textwidth}{!}{$\begin{aligned}
&\beta_{\lambda_2}^{ij} \Big|_{l^2} = 
l^2 \bigg[
 -\frac{1}{24} g^4 \left(N_c-1\right) \left(64 N_c-171\right) \lambda _{2}^{ij}
 +g^2 \left(\left(N_c-1\right) \lambda _{2}^{ik} \lambda _{2}^{jk}
 +2 \lambda _1 \left(N_c^2+N_c-2\right) \lambda _{2}^{ij}\right)
 \\&\quad
 -\frac{1}{2} \left(N_c+16\right) \lambda _{2}^{ik}\lambda _{2}^{jl} \lambda _{2}^{kl}
 -6 \lambda _1 \left(N_c+2\right) \lambda _{2}^{ik}\lambda _{2}^{jk}
 -\frac{1}{2} \lambda _{2}^{ij} \left((\lambda _{2}^{kl})^2
 +5 \lambda _1^2 \left(N_c+2\right)\right)
 -2 \lambda _{2}^{kl} \lambda _{2}^{lm} \lambda _{3}^{ijkm}
 \\&\quad
 -\lambda _{2}^{kl} \lambda _{3}^{ilmn} \lambda _{3}^{jkmn}
 +\left(\frac{1}{12}\lambda _{2}^{jk} \lambda _{3}^{ilmn} \lambda _{3}^{klmn}
 -2 \lambda _{2}^{jk} \lambda _{2}^{lm} \lambda _{3}^{iklm}
 + (i\leftrightarrow j)\right)
 \bigg]\,,
 \end{aligned}$}
\end{equation}
\begin{equation}
\resizebox{\textwidth}{!}{$\begin{aligned}
&\beta_{\lambda_3}^{ijkl} \Big|_{l^2} = 
l^2 \bigg[
 \left(2 g^2 \left(N_c-1\right) N_c \lambda _{2}^{ij}\lambda _{2}^{kl} 
+ \text{2 perms}\right)
 + \left(\frac{N_c}{4}\lambda _{2}^{im} \lambda _{3}^{jkln} \lambda _{2}^{mn} 
 + \frac{1}{12} \lambda_3^{ijkm}\lambda_3^{lnop}\lambda_3^{mnop}
    +\text{3 perms}\right)
\\&\qquad
  -  \left(
2N_c\, \lambda _{2}^{ij} \lambda _{2}^{km} \lambda _{2}^{lm} 
 + N_c\,\lambda _{2}^{im} \lambda _{2}^{jn} \lambda _{3}^{klmn} 
 + \lambda_3^{ijmn}\lambda_3^{kmop}\lambda_3^{lnop}
 +\text{5 perms}\right)
 \bigg]\,,
 \end{aligned}$}
\end{equation}
while the two-loop expressions that were left implicit in~\eqref{eq:dim5ScalarQCD} are
\begin{align}
 \frac{d\,c^i_{\varphi \phi^4}}{d\log\mu} \Big|_{l^2}
 &=l^2\bigg[
    -\frac{1}{3} \lambda _2 \left(30 \lambda _1+48 \lambda _2+5 \lambda _3\right) c_{\varphi^3\phi^2}^{ijj}
    \nn\\&\hspace{-15mm}\quad+c_{\varphi\phi^4}^i \bigg(\frac{1}{36} \left(-9 \lambda _2^2 \left(3 N_c+20 N_f+136\right)-48 \lambda _3 \lambda _2 \left(N_f+2\right)+\lambda _3^2 \left(N_f+2\right)\right)
        \nn\\&\hspace{-15mm}\qquad\qquad
        +\frac{1}{12} g^4 \left(-64 N_c^2+271 N_c-207\right)+2 g^2 \left(N_c-1\right) \left(2 \lambda _1 \left(N_c+5\right)+\lambda _2\right)
        \nn\\&\hspace{-15mm}\qquad\qquad
        -9 \lambda _1^2 \left(3 N_c+14\right)-12 \lambda _1 \lambda _2 \left(N_c+6\right)\bigg)
    \nn\\&\hspace{-15mm}\quad
    +c_{\varphi F^2}^i \bigg(g^6 \left(-19 N_c^2+70 N_c-51\right)+\frac{1}{6} g^4 \left(N_c-1\right) \left(2 \lambda _1 \left(64 N_c-207\right)
        +39 \lambda _2 \left(N_c-4\right)\right)
        \nn\\&\hspace{-15mm}\qquad\qquad
        -g^2 \left(N_c-1\right) \left(4 \lambda _1^2 \left(N_c+5\right)+\lambda _2 \lambda _1 \left(2-3 N_c\right)-\lambda _2^2\right)\bigg) \bigg]
        \nnl
 \frac{d\,c^{ijk}_{\varphi^3\phi^2}}{d\log\mu}\Big|_{l^2} 
 &=l^2\bigg[ \bigg\{
c_{\varphi\phi^4}^i \delta^{jk} \left(2 g^2 \left(N_c^2
 +N_c
 -2\right) \lambda _2 
 -\frac{1}{6} \left(N_c
 +2\right) \lambda _2 \left(30 \lambda _1
 +48 \lambda _2
 +5 \lambda _3\right) \right)
 \nn\\&\quad
 +c_{\varphi F^2}^i \delta^{jk} \bigg(\frac{1}{12} g^4 \left(N_c
 -1\right) \left(2 \left(64 N_c
 -171\right) \lambda _2
 +\left(13 N_c
 -36\right) \lambda _3\right) 
 \nn\\&\qquad\qquad
 -\frac{1}{2} g^2 \left(N_c
 -1\right) \left(\left(N_c
 +2\right) \lambda _1 \left(8 \lambda _2
 -\lambda _3\right)
 +\lambda _2 \left(\left(8
 -3 N_c\right) \lambda _2
 +\lambda _3\right)\right)\bigg)
 \nn\\&\quad
 -\frac{1}{18}  c_{\varphi^3\phi^2}^{ill} \delta^{jk} 
 \left(18 \lambda _2^2
 +48 \lambda _3 \lambda _2
 +13 \lambda _3^2\right) + \text{2 perms} \bigg\}
 \nn\\&\quad
 +c_{\varphi^3\phi^2}^{ijk}\bigg(\frac{1}{24} \left(
 -64 N_c^2
 +235 N_c
 -171\right) g^4
 +\left(N_c
 -1\right) \left(2 \left(N_c
 +2\right) \lambda _1
 +3 \lambda _2\right) g^2
 \nn\\&\qquad\qquad -\frac{5}{2} \left(N_c
 +2\right) \lambda _1^2
 -\frac{1}{4} \left(15 N_c
 +2 N_f
 +192\right) \lambda _2^2
 -\frac{1}{12} \left(3 N_f
 +38\right) \lambda _3^2
 \nn\\&\qquad\qquad-18 \left(N_c
 +2\right) \lambda _1 \lambda _2
 -2 \left(N_f
 +10\right) \lambda _2 \lambda _3\bigg) 
 -\frac{1}{2} c_{\varphi^5}^{ijkll} \lambda _2 \left(16 \lambda _2
 +5 \lambda _3\right) \bigg]\,.
 \label{eq:dim5ScalarQCDTwoLoop}
\end{align}
The anomalous dimensions of the EOM operators at dimension five are given by
{\allowdisplaybreaks
\begin{align}
\frac{d\,d^i_{\varphi \phi^2,1}}{d\log\mu} &= 
    -\ep d^i_{\varphi \phi^2,1}
    +l\bigg[
    d_{\varphi\phi^2,1}^i \left(\lambda _1 \left(N_c+2\right)-\frac{3}{2} g^2 \left(N_c-1\right)\right)
    +\lambda _2 d_{\varphi^3}^{ijj}\bigg]
    \nn\\&
    +l^2\bigg[-\frac{1}{2} \lambda _2 \left(N_c+2\right) c_{\varphi\phi^4}^i
    -\frac{\lambda _3 c_{\varphi^3\phi^2}^{ijj}}{6}-2 \lambda _2^2 d_{\varphi^3}^{ijj} 
    \nn\\&\quad
    +c_{\varphi F^2}^i \left(g^4 \left(-\frac{13 N_c^2}{4}+\frac{49 N_c}{4}-9\right)-\frac{3}{2} g^2 \left(N_c-1\right) \left(\lambda _1 \left(N_c+2\right)+\lambda _2\right)\right)
    \nn\\&\quad
    +d_{\varphi\phi^2,1}^i \bigg(\frac{1}{36} \left(-9 \lambda _2^2 \left(N_c+2 N_f\right)-90 \lambda _1^2 \left(N_c+2\right)-\lambda _3^2 \left(N_f+2\right)\right)
        \nn\\&\qquad\qquad \qquad
        +\frac{1}{24} g^4 \left(-64 N_c^2+235 N_c-171\right)+2 g^2 \lambda _1 \left(N_c^2+N_c-2\right)\bigg)\bigg]\,,
    \nn\\
    \frac{d\,d^i_{\varphi \phi^2,2}}{d\log\mu} &= 
    -\ep d^i_{\varphi \phi^2,2}
    +l\bigg[ 2 \lambda _2 d_{\varphi\phi^2,2}^i-\frac{3}{2} g^2 \left(N_c-1\right) c_{\varphi F^2}^i \bigg]
    \nn\\&\quad
    +l^2\bigg[
   -\frac{1}{2} \lambda _1 \left(N_c+2\right) c_{\varphi\phi^4}^i-\frac{3 g^4 \left(N_c^2-3 N_c+2\right) c_{\varphi F^2}^i}{8 \epsilon }
   \nn\\&\quad
   +c_{\varphi F^2}^i \left(\frac{1}{24} g^4 \left(-119 N_c^2+377 N_c-258\right)+\frac{1}{2} g^2 \lambda _2 \left(N_c-1\right)\right)
   \nn\\&\quad
   +d_ {\varphi \phi^2,2}^i\bigg(\frac{1}{36} \big(9 \lambda _2^2 \left(N_c-8\right)-36 \lambda _1 \lambda _2 \left(N_c+2\right)-12 \lambda _3 \lambda _2 \left(N_f+2\right)
        \nn\\&\qquad\qquad
        +\lambda _3^2 \left(N_f+2\right)\big)+\frac{1}{2} g^2 \lambda _2 \left(N_c-1\right)\bigg)-\frac{\lambda _2 c_{\varphi^3\phi^2}^{ijj}}{2}\bigg]\,,
        \nn\\
\frac{d\,d^{ijk}_{\varphi^3}}{d\log\mu} &= 
    -\ep d^{ijk}_{\varphi^3}
    +l\bigg[ 
    \frac{2}{3} \lambda _3 c_{\varphi^3}^{ijk}
    +\lambda _2 N_c d_{\varphi\phi^2,1}^i \delta^{jk}
    +\frac{1}{3} \lambda _3 c_{\varphi^3}^{ill} 
        \delta^{jk}
    \bigg]
    \nn\\&\quad
    +l^2\bigg[ 
    -\frac{3}{2} g^2 \lambda _2 \left(N_c-1\right) N_c c_{\varphi F^2}^i 
    \delta^{jk}
    -\frac{2}{9} \lambda _3^2 c_{\varphi^3}^{ill} \delta^{jk}
    \nn\\&\quad
    +d_{\varphi\phi^2,1}^i \delta^{jk} \left(2 g^2 \lambda _2 \left(N_c-1\right) N_c-2 \lambda _2^2 N_c\right)
    -\frac{1}{2} \lambda _2 N_c c_{\varphi^3\phi^2}^{ijk}
    \nn\\&\quad
    +\left(-\frac{3}{4} \lambda _2^2 N_c-\frac{1}{36} \lambda _3^2 \left(3 N_f+22\right)\right) c_{\varphi^3}^{ijk}-\frac{1}{6} \lambda _3 c_{\varphi^5}^{ijkll}\bigg]\,.
\end{align}
}

\bibliographystyle{apsrev4-1_title}
\bibliography{biblio}
\end{document}